\documentclass[%
 reprint,
 superscriptaddress,
 amsmath,amssymb,
 prb,
floatfix,
]{revtex4-1}

\usepackage{graphicx}
\usepackage{dcolumn}
\usepackage{bm}

\usepackage{subfigure}

\usepackage{graphicx}
\usepackage{placeins} 

\setcitestyle{super}

\usepackage{booktabs}
\usepackage{csquotes}

\usepackage{url}

\newcommand*{\balancecolsandclearpage}{%
 \close@column@grid
 \clearpage
 \twocolumngrid
}

\usepackage[usenames, dvipsnames, svgnames, table]{xcolor}

\renewcommand{\vec}[1]{\boldsymbol{#1}}
\newcommand{\ii}{\mathrm i}
\newcommand{\h}{\hspace{1pt}}

\begin{document}

\setlength{\intextsep}{10pt plus 2pt minus 2pt}

\preprint{APS/123-QED}

\title{Ab initio electronic structure and optical conductivity of bismuth tellurohalides}

\author{Sebastian Schwalbe}
 \email{schwalbe@physik.tu-freiberg.de}
\author{Ren\'{e} Wirnata}
\author{Ronald Starke}
\affiliation{%
 Institute of Theoretical Physics, TU Bergakademie Freiberg, Leipziger Str.~23, D-09596 Freiberg, Germany
}%

\author{Giulio A. H. Schober}
\affiliation{
 Institute for Theoretical Physics, Heidelberg University, Philosophenweg 19, D-69120 Heidelberg, Germany
}

\author{Jens Kortus}
\affiliation{%
 Institute of Theoretical Physics, TU Bergakademie Freiberg, Leipziger Str.~23, D-09596 Freiberg, Germany
}%

\date{\today}

\begin{abstract}
We investigate the electronic structure, dielectric and optical properties of bismuth tellurohalides BiTeX (X = I, Cl, Br) by means of all-electron density functional theory. In particular, 
we present the {\itshape ab initio} conductivities and dielectric tensors calculated over a wide frequency range, and compare our results with the recent measurements by \citet{Akrap}, \citet{Makhnev}, and \citet{Rusinov15}. We show how the low-frequency branch of the optical conductivity 
can be used to identify characteristic intra- and interband transitions between the Rashba spin-split bands in all three bismuth tellurohalides. We further calculate the refractive indices and dielectric constants,
which in turn are systematically compared to previous predictions and measurements. We expect that our quantitative analysis will contribute to the general assessment of bulk Rashba materials for their potential 
use in \mbox{spintronics devices.}
\end{abstract}

\pacs{Valid PACS appear here}
\maketitle


\section{\label{sec:motivation}Introduction}

The coupling of spin and orbital degrees of freedom lies at the heart of modern spintronics device concepts, 
which aim at an ultrafast and low-power-consumption information processing beyond the reach of present-day electronics.\cite{Sinova, Barnes}
Such a coupling is realized by the Rashba effect,\cite{Rashba60} which generally results from a large atomic spin-orbit interaction (SOI) and the lack of inversion symmetry. Traditionally, the Rashba spin splitting (RSS) has mainly been 
observed in two-dimensional systems like surfaces or interfaces between different materials.\cite{LaShell96, Nitta97, Ast07, 2012giant} By contrast, the recently observed {\it giant} RSS in the polar semiconductor BiTeI even turned out to be a {\itshape bulk} material property.\cite{2011giant, Crepaldi}
 In particular, this implies the possibility of observing optical transitions between the spin-split energy bands, \cite{Lee} an unconventional orbital paramagnetism \cite{Schober} as well as an enhanced magneto-optical response in the infrared regime.\cite{Demko} Furthermore, the theoretical prediction of a pressure-induced topological phase transition towards a noncentrosymmetric topological insulating phase of BiTeI \cite{BiTeITop} has led to several (as of yet still controversial) experimental investigations.\cite{Xi13, Tran14, Park15}

Shortly after the discovery of giant bulk RSS in the semicondutor BiTeI, general conditions for its appearance have been formulated,\cite{BiTeI_WIEN2k} and a number of related compounds have been investigated such as BiTeCl and BiTeBr. \cite{Eremeev, Sakano, Akrap} These materials have a band structure different from BiTeI, and in particular display a smaller RSS. Nevertheless, they have attracted much interest in materials science due to their unique electronic structures and properties. For example, topological surface states were predicted to appear in BiTeCl at ambient pressure.\cite{Chen} Furthermore, the optical properties and Raman spectra of BiTeBr and BiTeCl have been investigated experimentally by \citet{Akrap}. There, it turned out that the optical properties of these two compounds are very similar despite their different space groups ($P3m1$ for BiTeBr, $P6_3mc$ for BiTeCl).\cite{Akrap} Thus, the bulk Rashba materials BiTeX (X = I, Cl, Br) do not only realize a tabletop laboratory for investigating relativistic electron dynamics,\cite{Lee} but are also regarded as promising candidates \cite{Rashba12} for the future application in spintronics devices such as the Datta-Das spin transistor.\cite{Datta89, Koo09}

Despite this tremendous theoretical and experimental progress, much remains to be done to fully characterize the electronic structure and properties of the bismuth tellurohalides. For example, the band gap of BiTeBr and BiTeCl is a matter of ongoing discussion.\cite{Sakano, Sasagawa} Furthermore, in the important experimental work of \citet{Akrap} the question was raised of how the RSS in BiTeCl and BiTeBr influences the interband electronic transitions. In fact, the optical conductivity is one of the most fundamental physical quantities for characterizing the spin and orbital states of matter.
Already in the first studies of the relativistic electron dynamics in BiTeI,\cite{Lee} the optical spectra served as a fingerprint to identify transitions between the Rashba-split energy bands. In these early works,\cite{Lee}
selected elements of the optical conductivity tensor for BiTeI were calculated by applying the Kubo formula on top of an 18-band tight binding model constructed from an {\itshape ab initio} Hamiltonian using maximally localized Wannier functions.\cite{BiTeI_WIEN2k,TB_Wannier1,TB_Wannier2,TB_Wannier3,TB_Wannier4} More recently, the optical conductivity of BiTeX has been measured independently  by \citet{Makhnev} in a wide energy range up to $5$ eV. For an unambiguous deduction of the microscopic electronic structure and dynamics, it is therefore desirable to systematically calculate the optical conductivity of the bismuth tellurohalides from first principles. 
Generally, the optical conductivity can also be used for the calculation
of the dielectric tensor and hence the refractive index, which has indeed been done for the case of BiTeI in the work of \citet{Rusinov15}.

In this article, we resume this line of research. In particular, we present the entire optical conductivity and dielectric tensors of the bismuth tellurohalides (BiTeI,~BiTeCl,~BiTeBr)
calculated {\itshape ab initio} from density functional theory (DFT).
In order to prove in the first place that the bismuth tellurohalides can be treated reliably within DFT, we will first compute several completely relaxed electronic and structural properties,
which we will later compare to experimental data. Then, we will identify those elements which mainly contribute to the Rashba effect
in the atom-species resolved projected density of states (PDOS). Furthermore, we will show that the electron localization function (ELF) \cite{ELF}
displays a layered structure. After that, we will report the frequency-dependent optical conductivity of BiTeI for different values of the Fermi energy. This will in turn enable us to identify the Rashba-specific peaks in the optical conductivity, which correpond to intra- and interband transitions between the Rashba-split bands. Finally, we will provide a comparison of our {\itshape ab initio} results for the optical conductivity spectra, dielectric constants and refractive indices  with the recent experiments performed by \citet{Akrap}, \citet{Makhnev} and \citet{Rusinov15}.


\section{\label{sec:theory}Theoretical details}

As mentioned in the introduction, we will report calculations of the dielectric tensor and the refractive index of BiTeX (X = I, Cl, Br).
The basic quantity computed by the ELK code is, however, the conductivity, and thus we have to clarify its relation to the aforementioned material properties (for details see \citet{Giuliani}). For simplicity, we restrict ourselves to scalar relations of wavevector-independent quantities.
As a matter of principle, the (direct) conductivity relates the induced current to the external electric field by means of
$
\vec j_{\rm ind}=\sigma\vec E_{\rm ext}.
$
By contrast, the proper conductivity relates the induced current to the total electric field,
$
\vec E_{\rm tot}=\vec E_{\rm ext} + \vec E_{\rm ind},
$
by means of $\vec j_{\rm ind} = \widetilde\sigma \vec E_{\rm tot}$.
Finally, the dielectric function mediates between the external and the total electric field in the sense of
$
\vec E_{\rm ext}=\varepsilon_{\rm r} \h \vec E_{\rm tot}.
$
These quantities are related by the well-known equations
\begin{align}
\varepsilon_{\rm r}^{-1}(\omega) &= 1+\frac{\sigma(\omega)}{\ii\omega\varepsilon_0}\,,\label{eq_notuse}\\[3pt]
\varepsilon_{\rm r}(\omega) &= 1-\frac{\widetilde\sigma(\omega)}{\ii\omega\varepsilon_0} \label{eq_use}\,.
\end{align}
However, although our calculations of the conductivity are based on the Kubo formalism and hence yield the direct conductivity
, we actually use Eq.~\eqref{eq_use} to perform the transition to the dielectric function.
In other words, we interpret the conductivity calculated by the ELK code as the {\it proper} conductivity.
This, of course, requires a certain justification, for which we will provide in the following. For this purpose, we start from the standard relations
\begin{align}
\varepsilon_{\rm r}^{-1}(\omega) &= 1+v\h\chi(\omega)\,, \label{eq_worse}\\[3pt]
\varepsilon_{\rm r}(\omega) &= 1-v\h\widetilde\chi(\omega)\,, \label{eq_better}
\end{align}
between the dielectric function and the (direct and proper) density response functions respectively defined by
$\chi = \delta\rho_{\rm ind}/\delta\varphi_{\rm ext}$ and $\widetilde\chi = \delta\rho_{\rm ind}/\delta\varphi_{\rm tot}$, where $\varphi$ is the scalar potential while $v$ denotes the Coulomb interaction kernel.
Using the functional chain rule, one shows directly that the direct density response function is related to its proper counterpart
by the self-consistent equation
\begin{equation}
 \chi = \widetilde\chi + \widetilde\chi \h v \h \chi\,.
\end{equation}
Suppose now that we consider a many-body system and we are given the density response function $\chi_0$ in a noninteracting approximation,
as it is indeed the case for DFT.
In that case, $\chi_0$ describes the density response function under the assumption that the constituents of the system do not interact with each other.
We now want to approximate the true, i.e. interacting response function $\chi$ of the system by means of its noninteracting counterpart $\chi_0$.
It is plausible that we can do this, if we simply take $\chi_0$ as the response to both the external field {\it and} the induced field generated by the electrons
themselves. This approach takes the interactions into account by simply assuming that the electrons ``feel'' their own induced field in addition to the external field.
Concretely, this means to re-interpret $\chi_0$ as an approximation for the {\it proper} response function, such that the desired
approximation for the interacting response function is given by
\begin{equation}
 \chi=\chi_0+\chi_0 \h v \h \chi\,.
\end{equation}
This equation constitutes the {\it random phase approximation} (see Chap.~5.3.1.1 in \citet{Giuliani}). Thus, in order to calculate
the dielectric function $\varepsilon_{\rm r}(\omega)$ from $\chi_0(\omega)$ one has to use Eq.~\eqref{eq_better} rather than Eq.~\eqref{eq_worse}.
Correspondingly, as both in the direct and proper case the conductivity and density response function are related by
\begin{equation}
\sigma(\omega)=\ii\omega \h \varepsilon_0 \h v \h \chi(\omega) \,,
\end{equation}
the dielectric function should be calculated by means of Eq.~\eqref{eq_use} rather than Eq.~\eqref{eq_notuse}. This concludes our discussion of the relation between the conductivity and the dielectric function. Finally, the frequency-dependent refractive index
can be directly evaluated as
$n(\omega)=\sqrt{\varepsilon_{\rm r}(\omega)}$. In the case of a wavevector dependence, the relation between the dielectric tensor and the refractive index may become more complicated,\cite{Refr} which is however not considered in this article.

\section{\label{sec:computation}Computational details}

Our calculations have been performed with the ELK \cite{ELK} code, which relies on a full potential (FP), linear augmented plane-wave (LAPW) basis.\cite{LAPW1,LAPW2,LAPW3,LAPW4}
Concretely, we have used a dense $k$-grid of $20\times 20 \times 20$ $k$-points and a cutoff value of $R_{\rm MT}K_{\rm max} = 7$. For the electronic structure calculations we have employed the PBE-GGA \cite{pbe} exchange-correlation functional. 
Finally, in order to obtain a Rashba splitting in the band structure, spin-polarization and SOI have been taken into account.\cite{SpinSplitting1,SpinSplitting2}

\begin{figure}[h]
 \centering
 \includegraphics[width=1.0\linewidth]{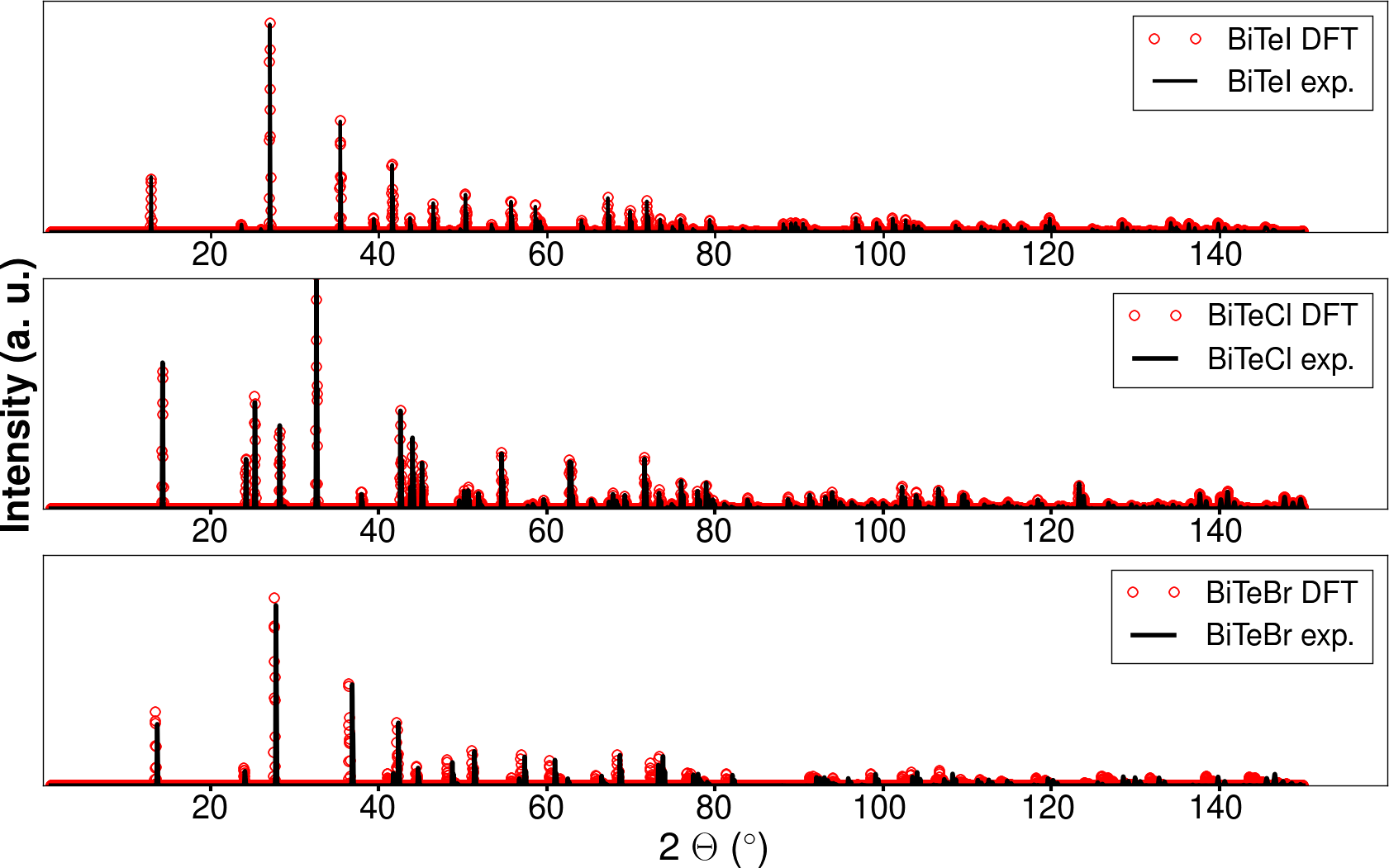}
 \caption{Calculated powder diffraction pattern for BiTeX (X = I,~Cl,~Br). The black lines in each subplot visualize the experimental structures, whereas the red dots correspond to the {\itshape ab initio} unit cells, where atomic positions and cell parameters have been relaxed.}
 \label{PDP_FIG}
\end{figure}

We have used experimental lattice structures \cite{XRAY_STRUCTURES} as an input to
subsequently optimize the lattice constants and volumes to a minimal absolute force value of $0.5 \times 10^{-3}$ Hartree/Bohr.
In order to ascertain that our relaxed crystal structures really correspond to the experiment, we have also calculated 
the corresponding powder diffraction patterns (PDP) for experimental as well as theoretically optimized structures (see Fig.~\ref{PDP_FIG}; cf.~also \citet{BiTeI_WIEN2k}). We have verified that the optimized DFT structures produce a similar powder diffraction pattern as compared to their experimental counterparts. 
For these computations we have used the open source FullProf program suite.\cite{Fullprof}
In particular, we have meticulously simulated X-ray patterns in the Bragg-Brentano geometry.

As in the work of \citet{Lee}, charge carrier concentrations have been taken into account for the calculation of the 
optical conductivity by adjusting the Fermi level manually,
for which purpose the ELK code had to be modified appropriately (see Fig.~\ref{BiTeI_Ef}(b)).
We note that the Fermi level had to be manipulated only for this concrete calculation.
The resulting effect has then been used to show the Rashba-specific features in the entries of the optical conductivity tensor.

Finally, the visualizations of the ELF have been plotted with VESTA, \cite{momma_vesta_2011} using the calculated ELF with an iso-level of 0.55 (min=0, max=0.80). 

\section{\label{sec:results}Results and discussion}

\begin{figure}[t!]
 \centering
 \subfigure[~BiTeI]{\includegraphics[width=\linewidth]{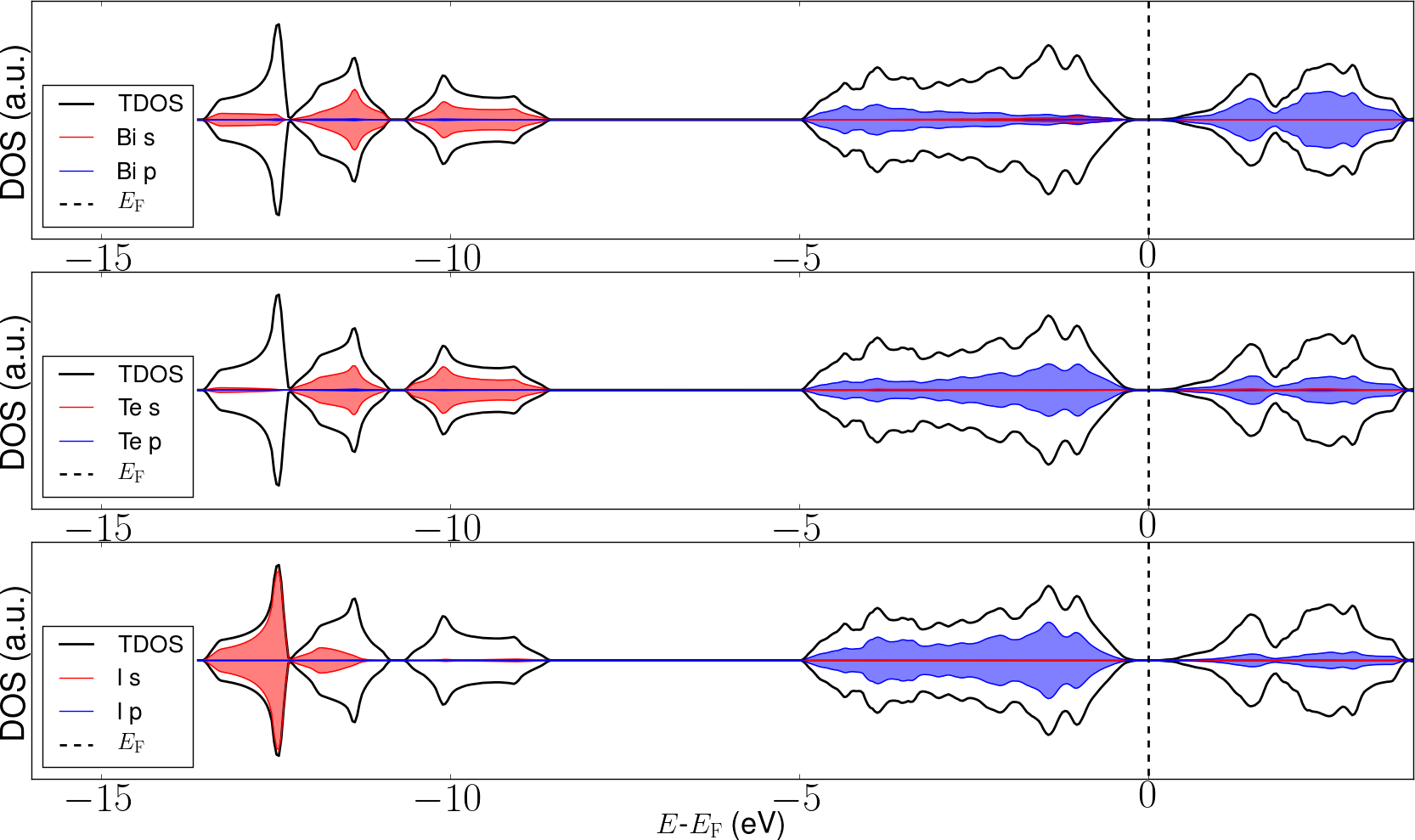}}
 \subfigure[~BiTeCl]{\includegraphics[width=\linewidth]{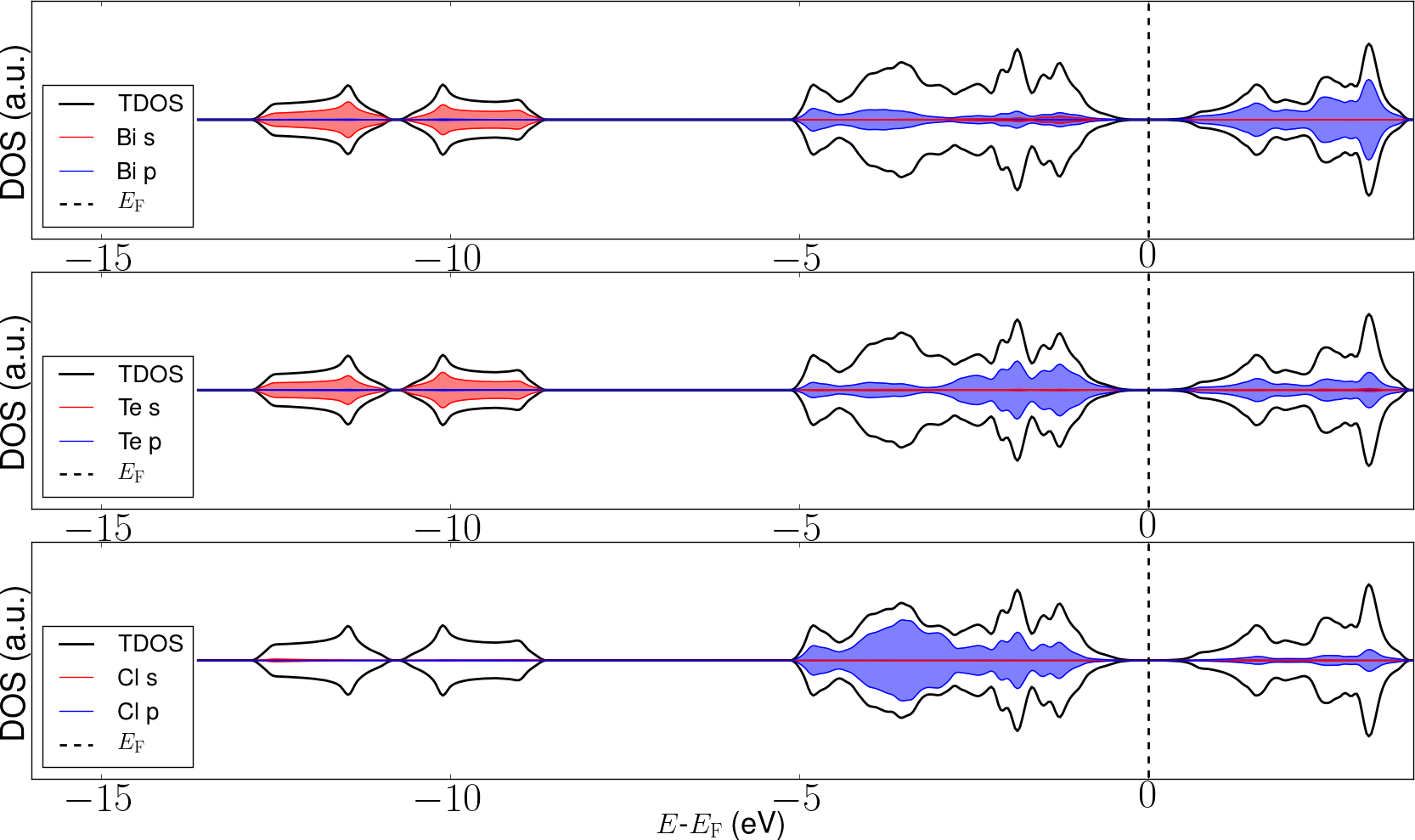}}
 \subfigure[~BiTeBr]{\includegraphics[width=\linewidth]{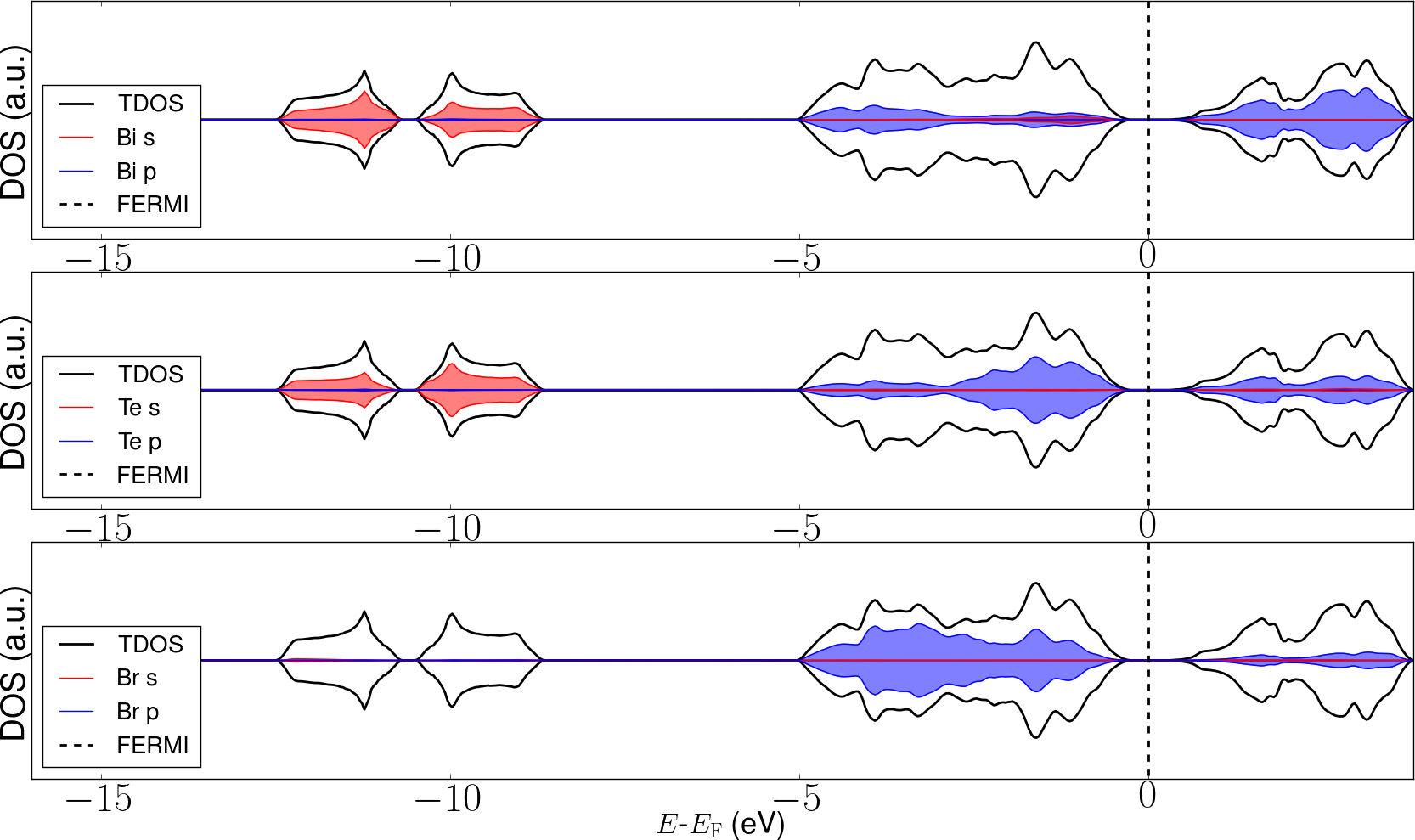}} 
 \caption{Total density of states (TDOS) and its projections (PDOS) to s- and p-states of the constituent atoms for each halide. The corresponding Fermi energies $E_{\text{F}}$ and band gaps $E_{\text{G}}$ are given in Table \ref{tab_E_F}. All Fermi levels are exclusively determined by p-states, where the tellurium and halogen atoms show the largest share. The underlying s-states contribute to the DOS not until 10 meV below the Fermi level. In contrast, unoccupied states are mainly determined by p-states of Bi atoms.}
 \label{BiTeX_PDOS}
\end{figure}

For all bismuth tellurohalides, the projected density of states (PDOS) shows that the Fermi level is determined by the p-states of all atom species, 
whereas the unoccupied states are dominated by the bismuth p-states (see Fig.~\ref{BiTeX_PDOS}). Consequently, the s-states are far below the Fermi level (approximately 10 eV). Neither do d-states play any r\^{o}le in the DOS around the Fermi level. 
By contrast, the states in the valence band are dominated by the halide orbitals. 
Correspondingly, the reason for the differences between the bismuth tellurohalides structures 
can be ascribed to the contribution of the respective halogen atoms to the valence states. While for BiTeCl and BiTeBr three clearly separated s-bands show up, in the case of BiTeI the corresponding lowest s-bands \mbox{overlap.}

We have also calculated the electronic band gap $E_{\text{G},\text{DFT}}$, which we compare to the available experimental data in Table \ref{tab_E_F}. 
While our predicted band gap is quite generally in remarkable agreement with the experiment for BiTeI (see \citet{2011giant}), previous theoretical predictions for BiTeCl and BiTeBr differ from it.\cite{2011giant,Akrap,Rusinov13,BiTeBr_optics} However, our calculations for BiTeCl and BiTeBr yield smaller band gaps as compared to the experimental values. Nevertheless, the overall agreement is satisfactory in our calculations (compare \citet{Rusinov13}, \citet{BiTeBr_optics}, and \citet{Guo16}). 

\begin{table}[h!]
\centering
\caption{Fermi energies $E_{\text{F}}$ and band gaps $E_{\text{G}}$ for the three bismuth tellurohalides BiTeX (X = I, Cl, Br)}
\begin{tabular}{c|c|c|c}
 \toprule[1pt]
 System & $E_{\text{F}}$ (eV) & $E_{\text{G},\text{DFT}}$ (eV) & $E_{\text{G},\text{exp}}$ (eV) \\
 \midrule[0.5pt]
 BiTeI  & 4.58 & 0.37 & 0.38\cite{2011giant}\\ 
 BiTeCl & 4.26 & 0.50 & 0.77\cite{Akrap}  \\
 BiTeBr & 4.03 & 0.55 & 0.62\cite{Akrap}\\ 
 \bottomrule[1pt]
\end{tabular}
\label{tab_E_F}
\end{table}

\begin{figure}[ht!]
 \centering
 \subfigure[~BiTeI]{\includegraphics[width=0.45\linewidth]{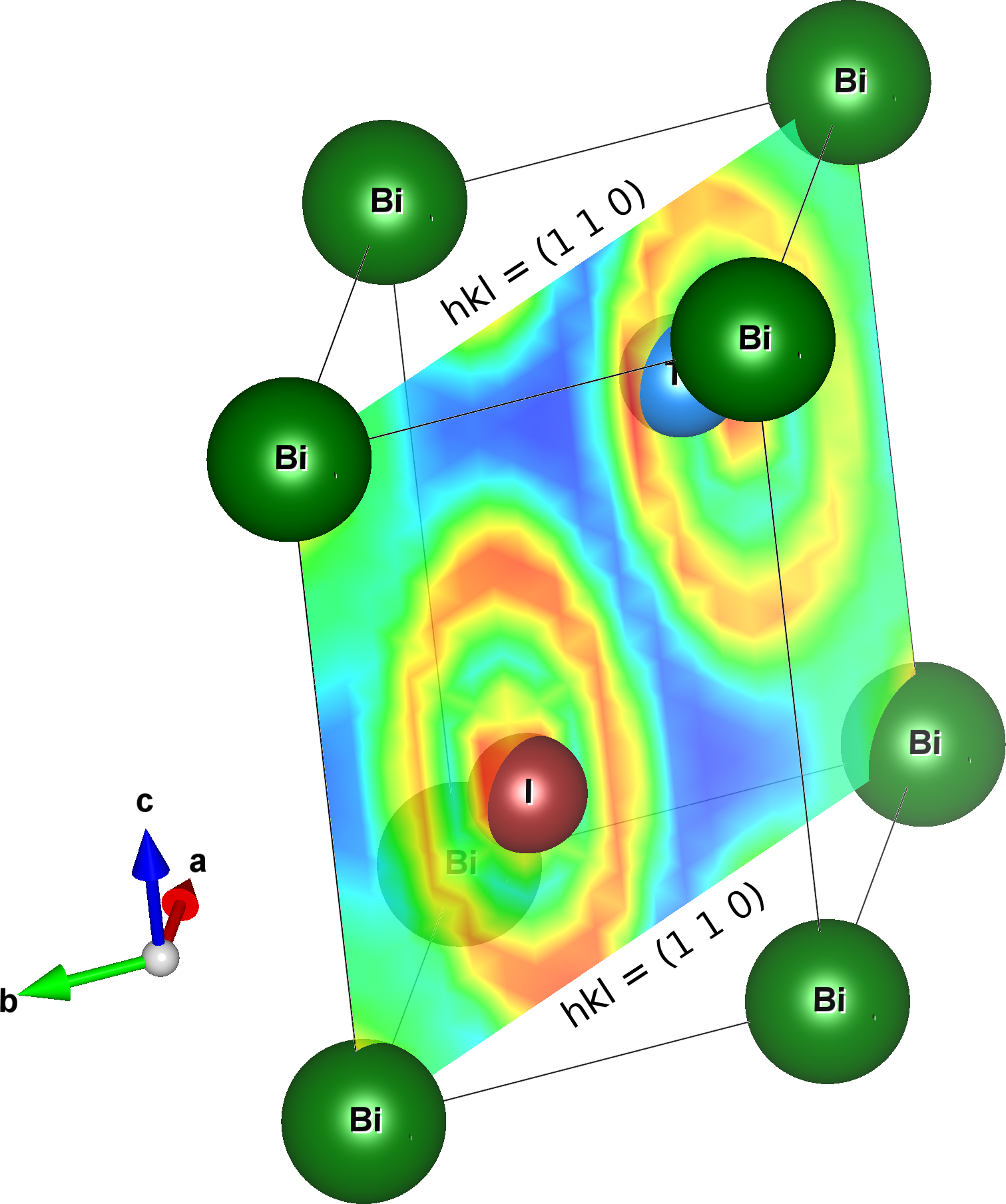}
  \hspace{2em} \includegraphics[width=0.05\linewidth]{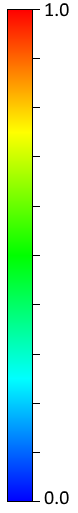}}
 \subfigure[~BiTeCl]{\includegraphics[width=0.45\linewidth]{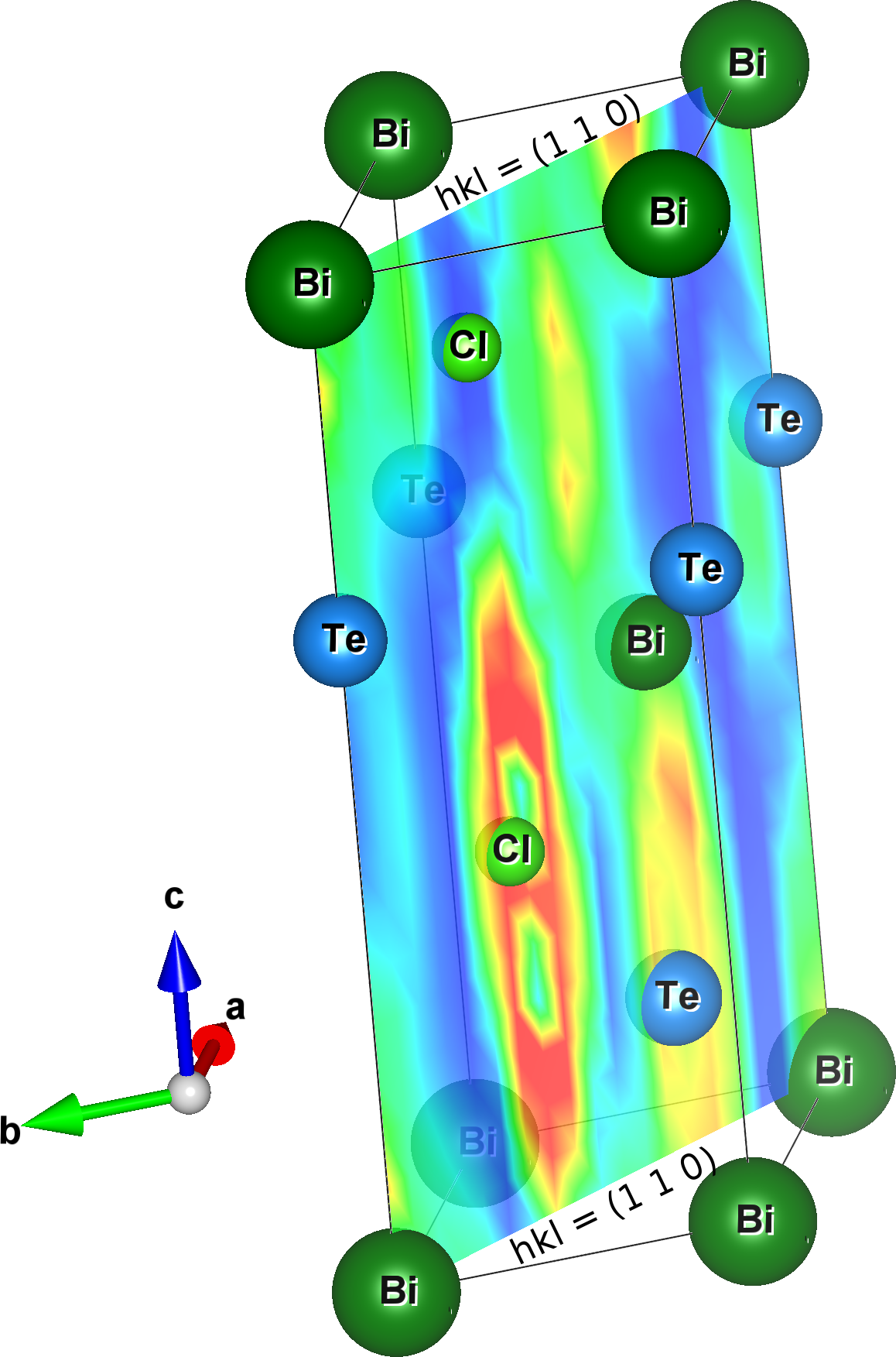}
  \hspace{2em} \includegraphics[width=0.05\linewidth]{ELF_scale.png}}
 \subfigure[~BiTeBr]{\includegraphics[width=0.45\linewidth]{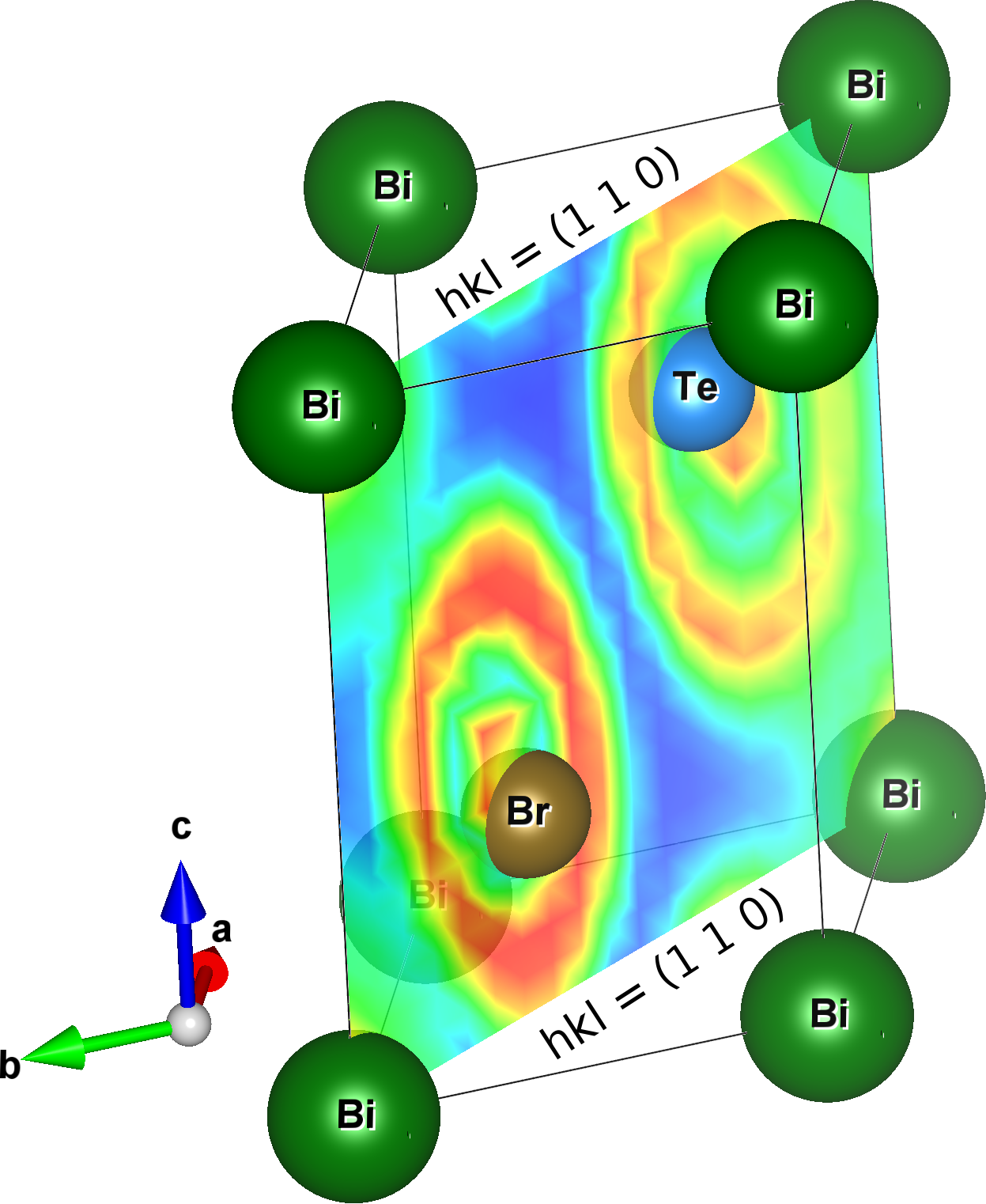}
 \hspace{2em} \includegraphics[width=0.05\linewidth]{ELF_scale.png}}
 \caption{Electron localization function (ELF) for the three bismuth tellurohalides BiTeX (X = I,~Cl,~Br). 
 Red parts correspond to areas with highly localized electron density, whereas blue parts symbolize less strong localization. 
 Two layers of higher localized density are formed, one centered at the tellurium atom and the other one centered at the halogen atom.}
 \label{BiTeX_ELF}
\end{figure}

Turning to the ELF (see Fig.~\ref{BiTeX_ELF}), we first observe that in the case of BiTeI the electrons are localized in two layers (\enquote{2d-electron gas}), 
one layer around the Te atom and the other layer around the I atom. This 2d-electron gas can also be observed in BiTeCl and BiTeBr. In the case of BiTeBr, 
the electron density is mainly localized at the Br and the Te atoms in the form of two seperate layers. Similarly, the localization layers 
of BiTeCl are centered at the Cl and the Te atom. However, all bismuth tellurohalides have in common that one electron localization layer is formed by the Te atom and the other by the halogen atom. 
This layered structure is well known for the case of BiTeI,\cite{Kilic} but to our knowledge not generally discussed 
for the whole class of bismuth tellurohalides.

\begin{figure}[hb!]
 \centering
 \subfigure[~BiTeI: low-frequency]{\includegraphics[width=0.329\linewidth]{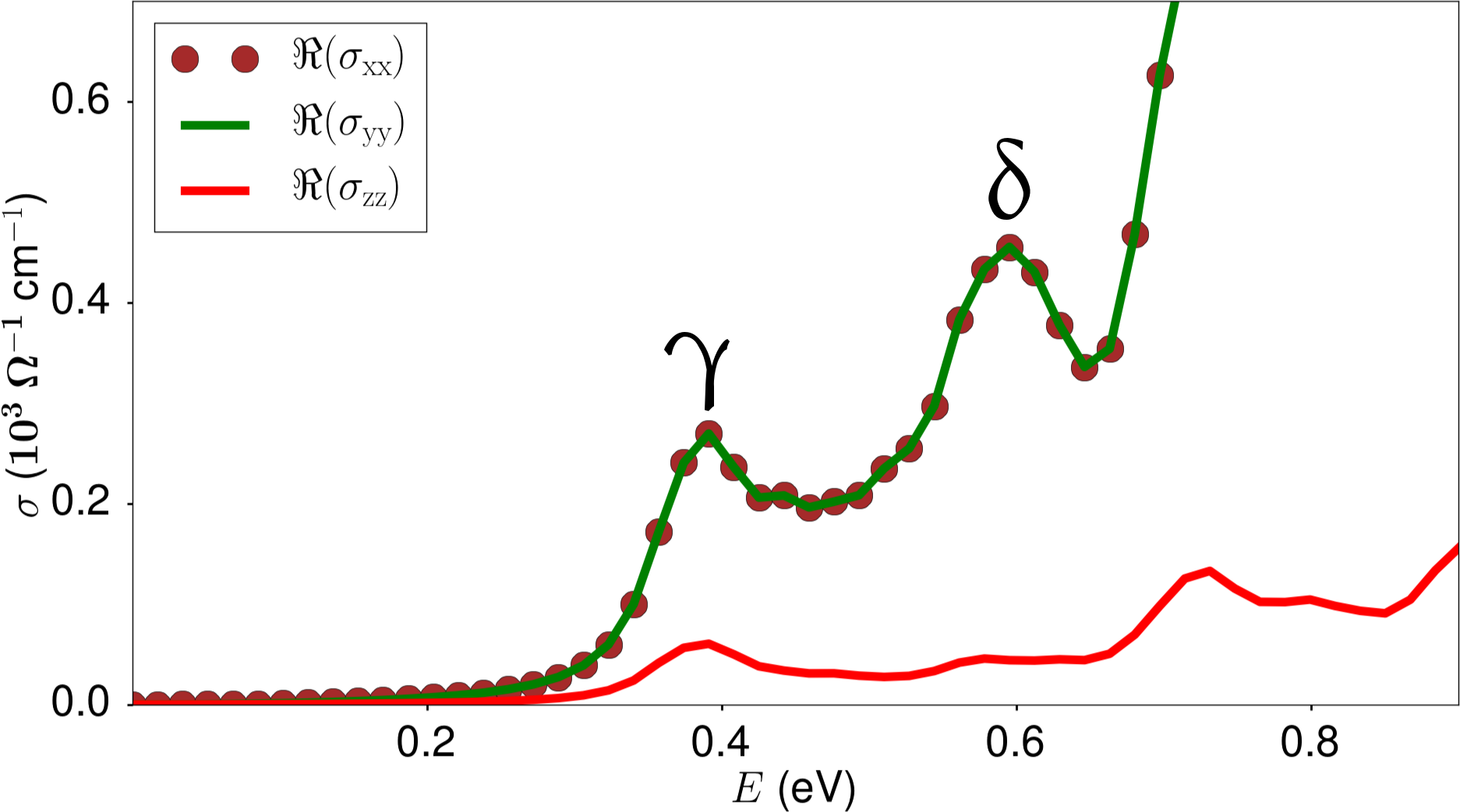}} 
 \subfigure[~BiTeCl: low-frequency]{\includegraphics[width=0.329\linewidth]{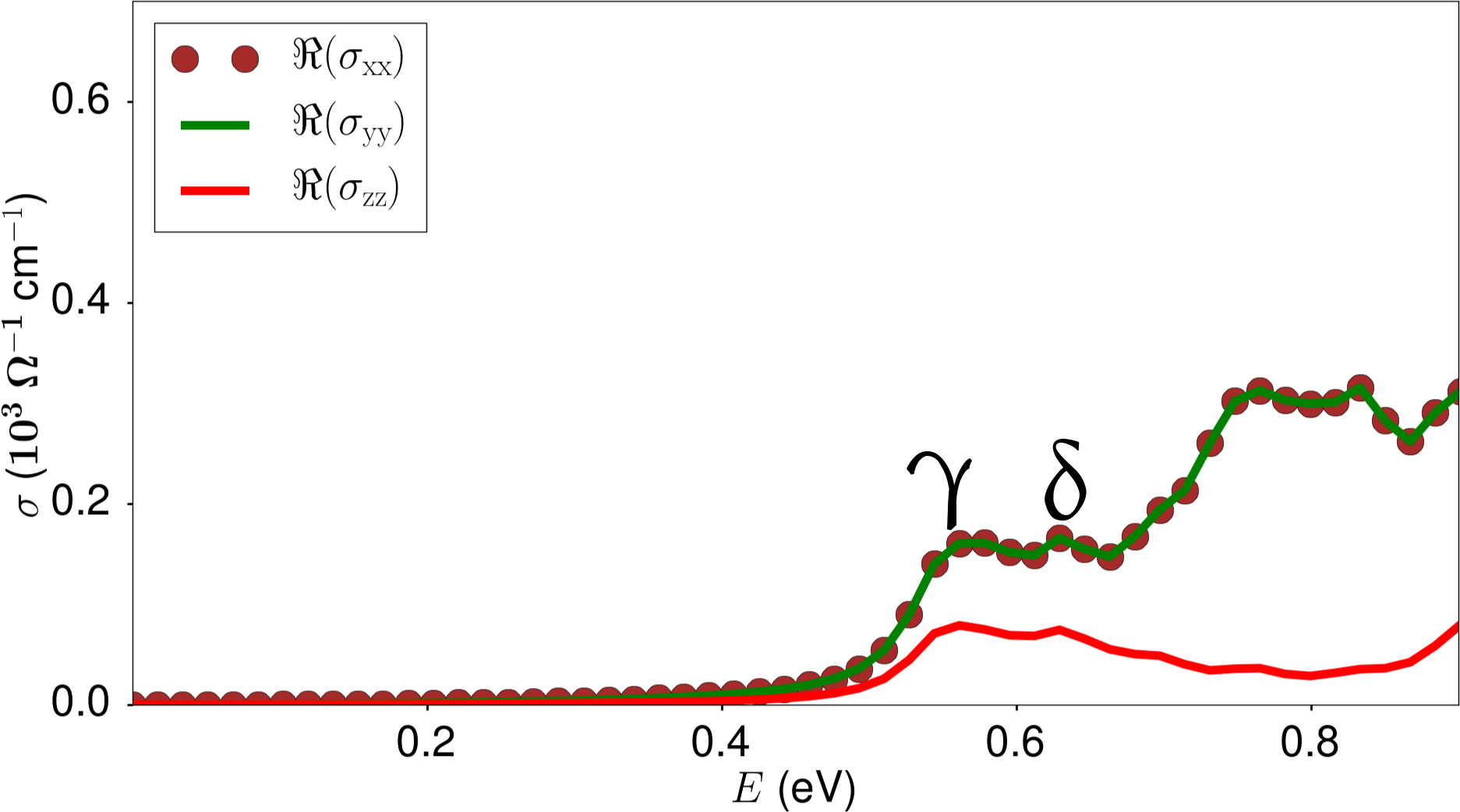}} 
 \subfigure[~BiTeBr: low frequency]{\includegraphics[width=0.329\linewidth]{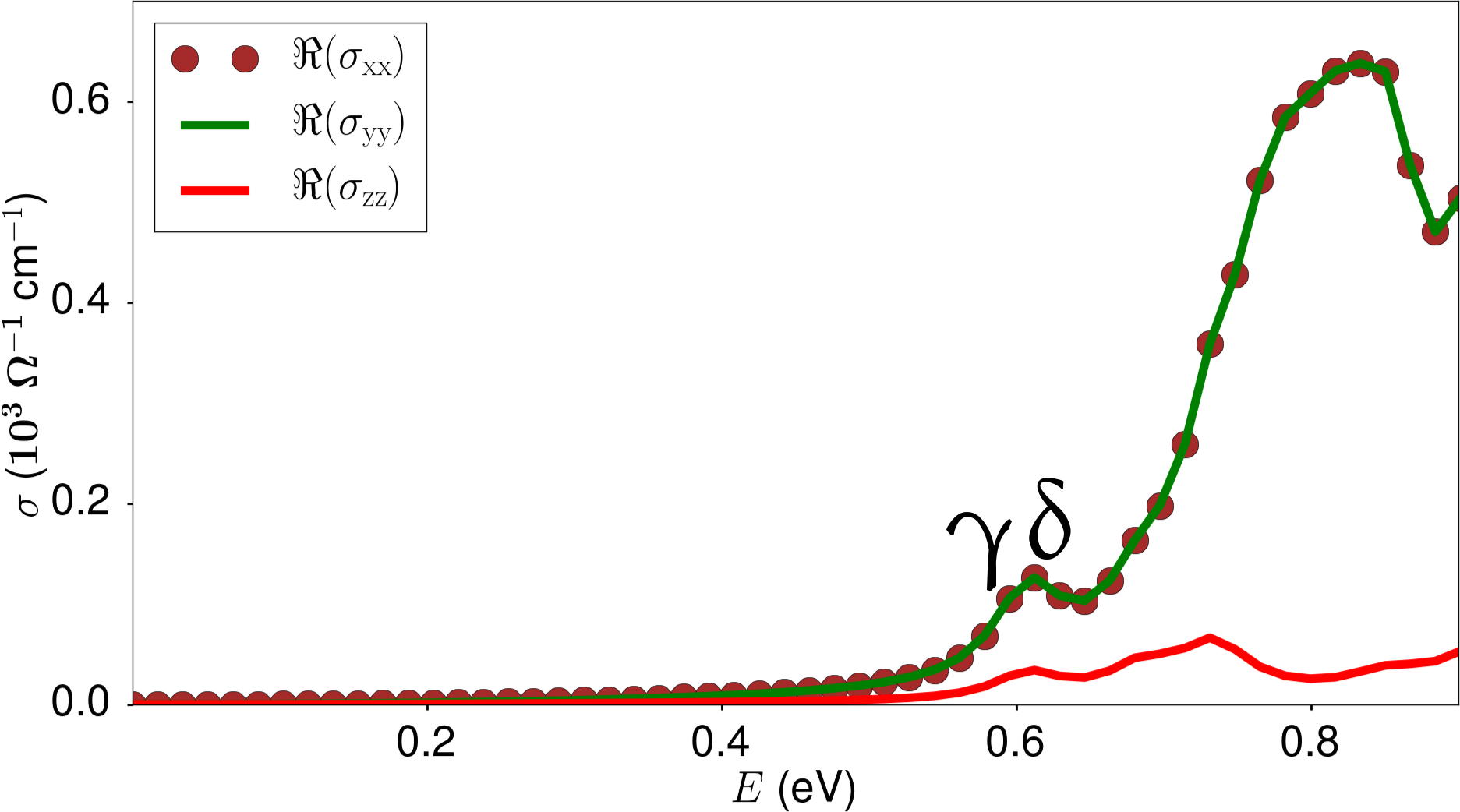}}
 \subfigure[~BiTeI: high-frequency]{\includegraphics[width=0.329\linewidth]{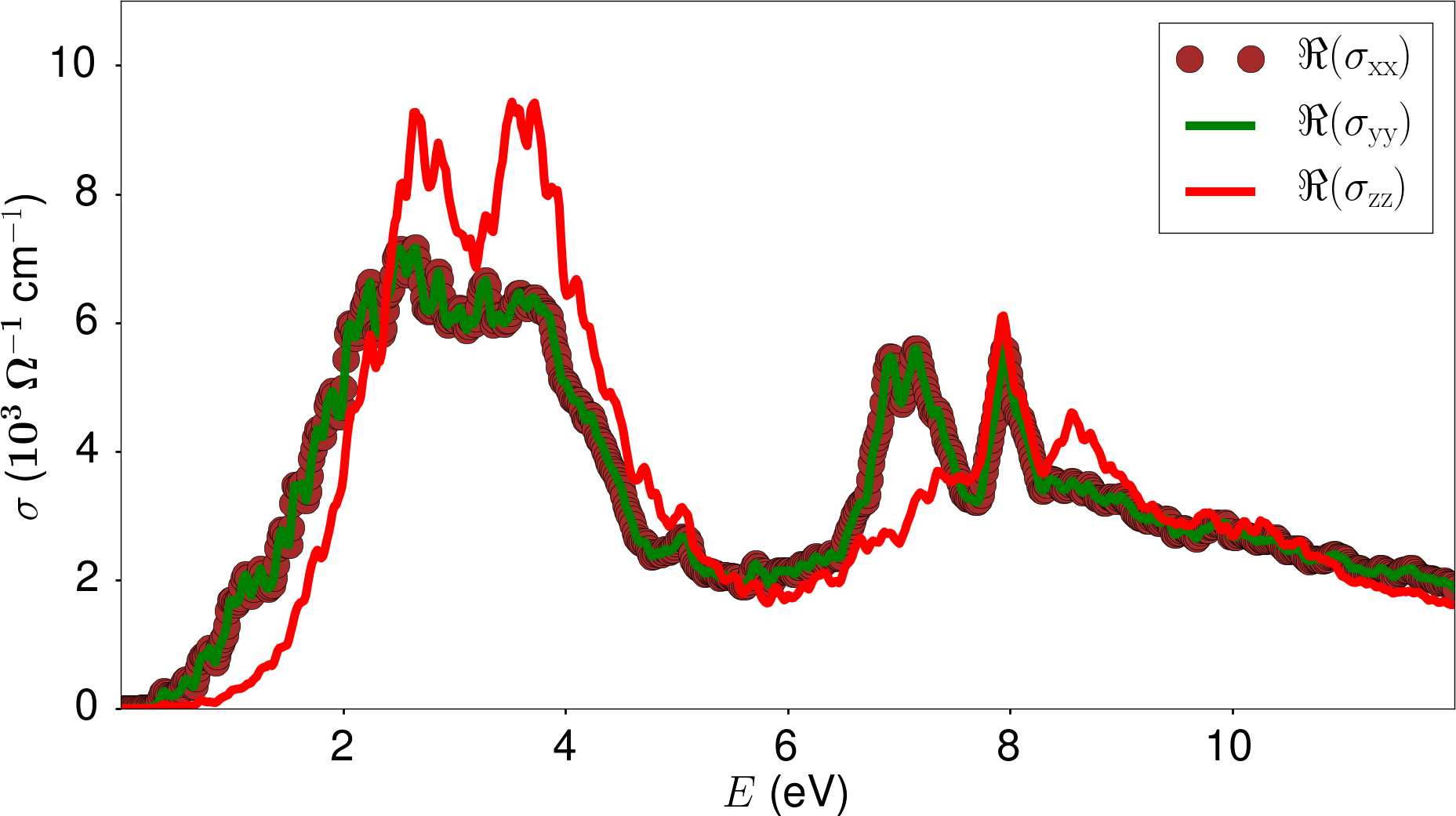}}
 \subfigure[~BiTeCl: high-frequency]{\includegraphics[width=0.329\linewidth]{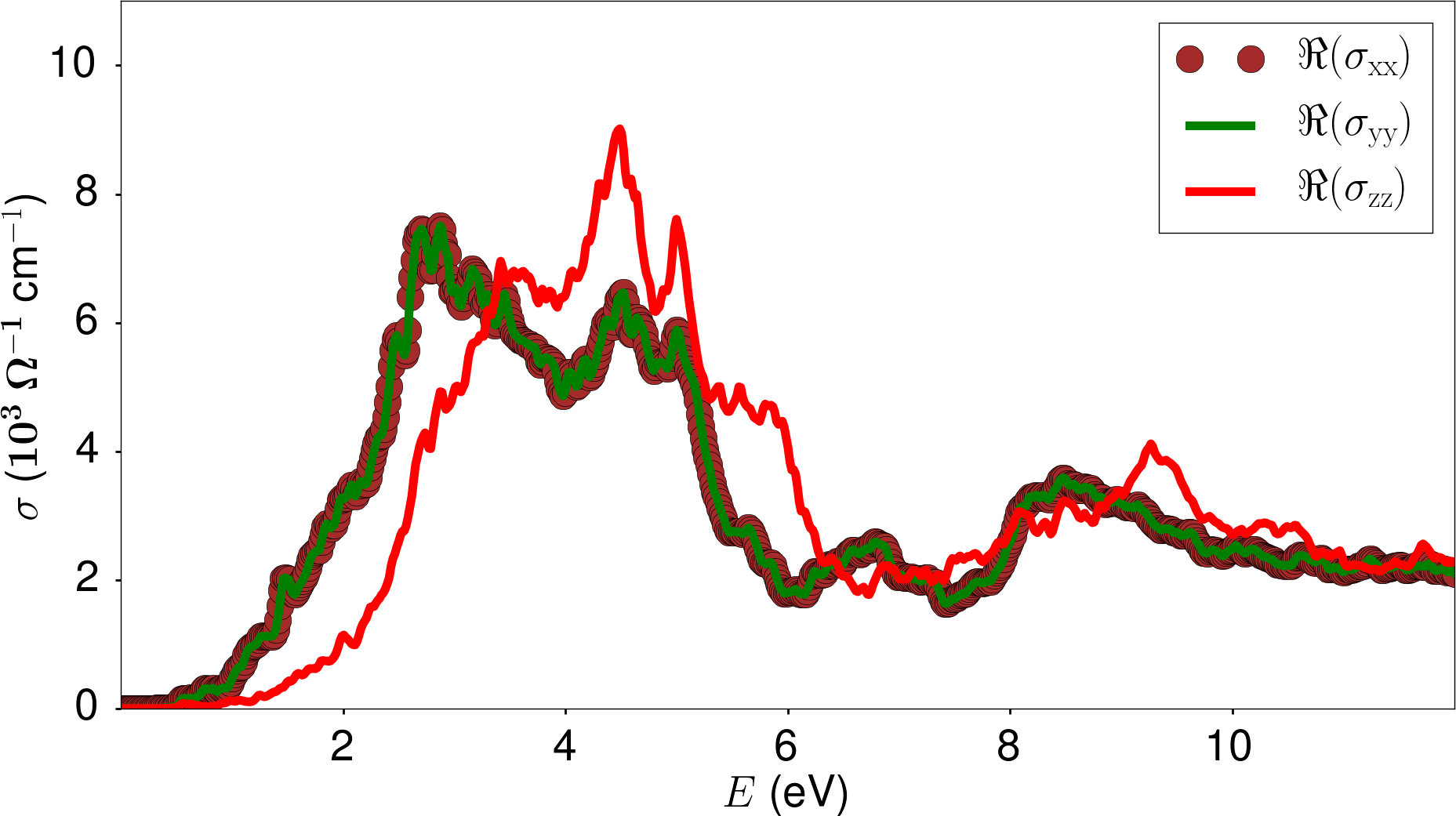}}
 \subfigure[~BiTeBr: high-frequency]{\includegraphics[width=0.329\linewidth]{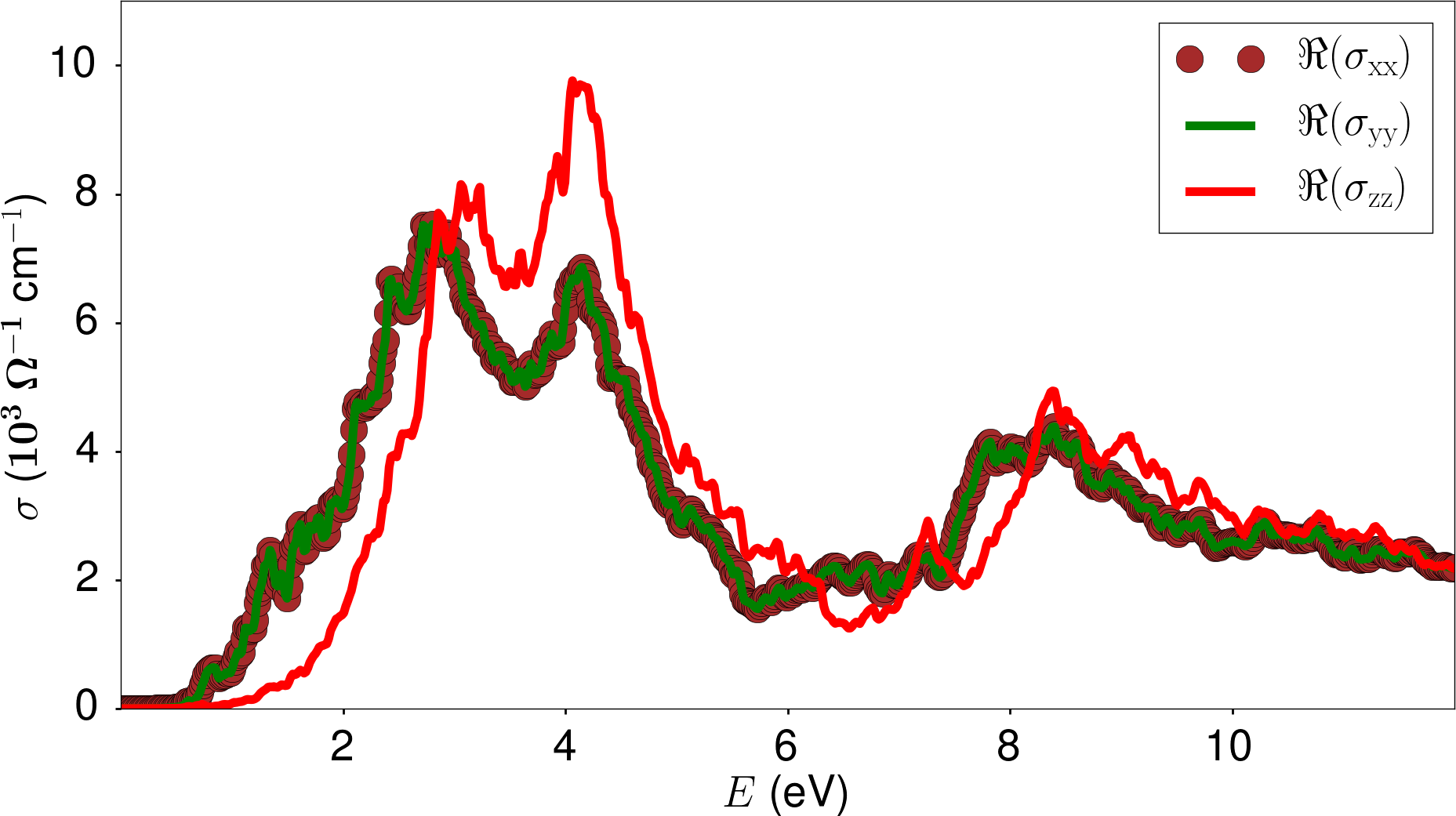}}
 \caption{Real parts of the optical conductivities for BiTeX (X = I,~Cl,~Br). The low-frequency regions are plotted separately in (a), (b) and (c). The 
 characteristic energies marked with $\gamma$ and $\delta$ correspond to interband transitions (cf.~Fig.~\ref{BiTeI_Ef}(b)).} 
 \label{BiTeX_sigma}
\end{figure}

Moreover, we have calculated the optical conductivity tensor of BiTeX (X = I, Cl, Br) in a wide energy range up to 12 eV.
In comparing its diagonal elements, one observes similarly shaped spectra with coinciding orders of magnitude
in $\sigma_{xx}$ and $\sigma_{yy}$, whereas $\sigma_{zz}$ displays a completely different behaviour. This is in fact the case
for all bismuth tellurohalides (see Fig.~\ref{BiTeX_sigma}). Interestingly, the lower frequency branches of the conductivity are generally peaked in the $(110)$-plane. The high-frequency spectra of BiTeCl and BiTeBr are very similar despite their different point groups, which is in accord with the experimental findings of \citet{Akrap}. 
Furthermore, our calculated optical conductivity spectrum of BiTeI agrees well with the measurements of \citet{Makhnev} (see Fig.~3 there) over a wide frequency range. 

\begin{table}[h!!!!!!]
\centering
\caption{Relative dielectric constants $\varepsilon_{{\rm r},ii} (\omega = 0)$ and refractive indices $n (\omega = 0 )$ for BiTeX (X = I, Cl, Br).}
\begin{tabular}{c|c|c|c|c}
 \toprule[1pt]
 System & $\varepsilon_{{\rm r},xx} (\omega = 0)$& $\varepsilon_{{\rm r},yy} (\omega = 0)$ &  $\varepsilon_{{\rm r},zz} (\omega = 0)$ & $n (\omega = 0 )$ \\ 
 \midrule[0.5pt]
 BiTeI  & 21.64 & 21.52 & 15.54 & 4.65\\ 
 BiTeCl & 15.11 & 15.08 & 10.38 & 3.89 \\
 BiTeBr & 16.59 & 16.60 & 11.64 & 4.07 \\ 
 \bottomrule[1pt]
\end{tabular}
\label{tab_n}
\end{table}

\FloatBarrier

\onecolumngrid


\begin{figure}[ht!]
 \centering
 \subfigure[~BiTeI]{\includegraphics[width=0.329\linewidth]{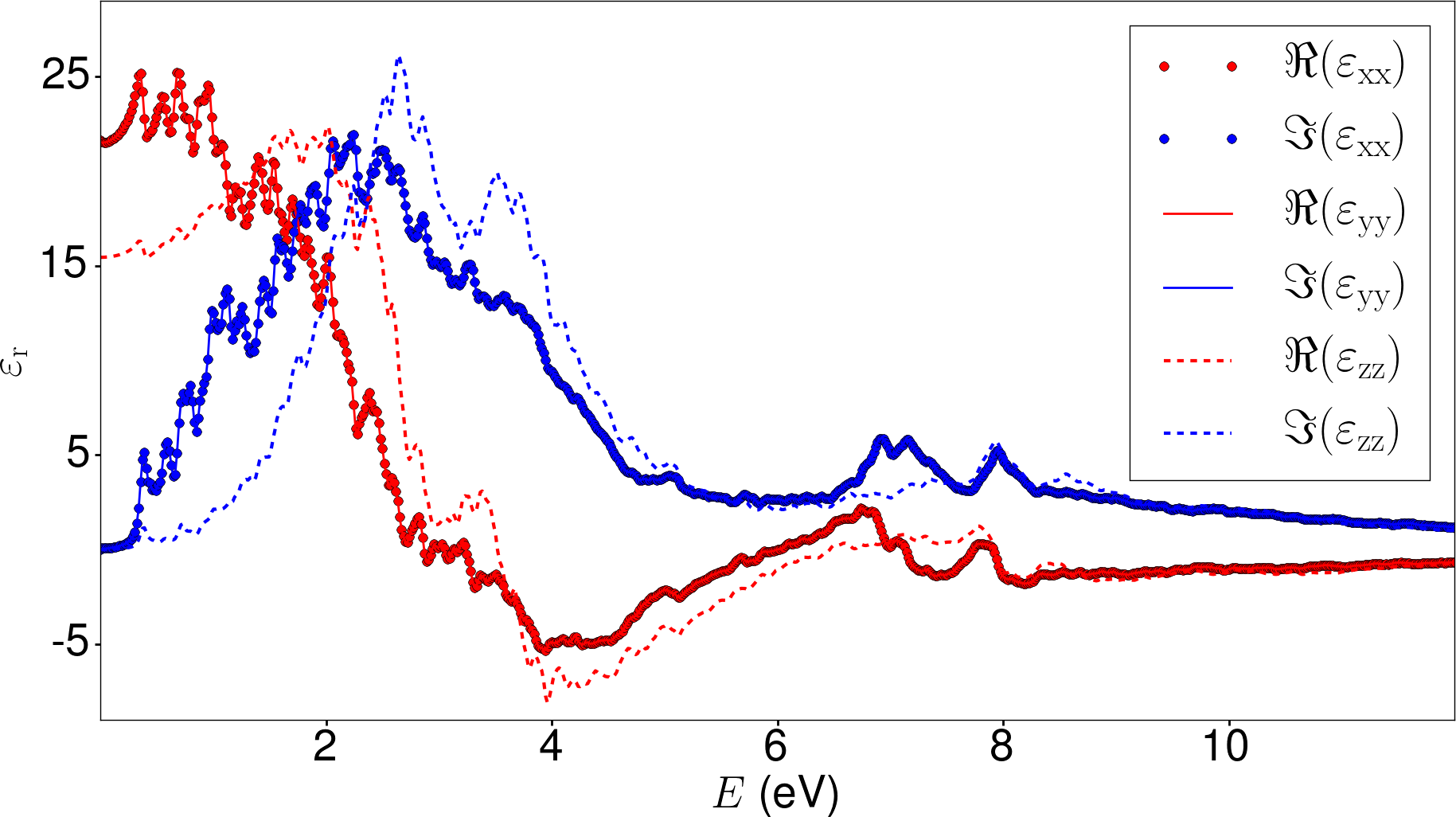}}
 \subfigure[~BiTeCl]{\includegraphics[width=0.329\linewidth]{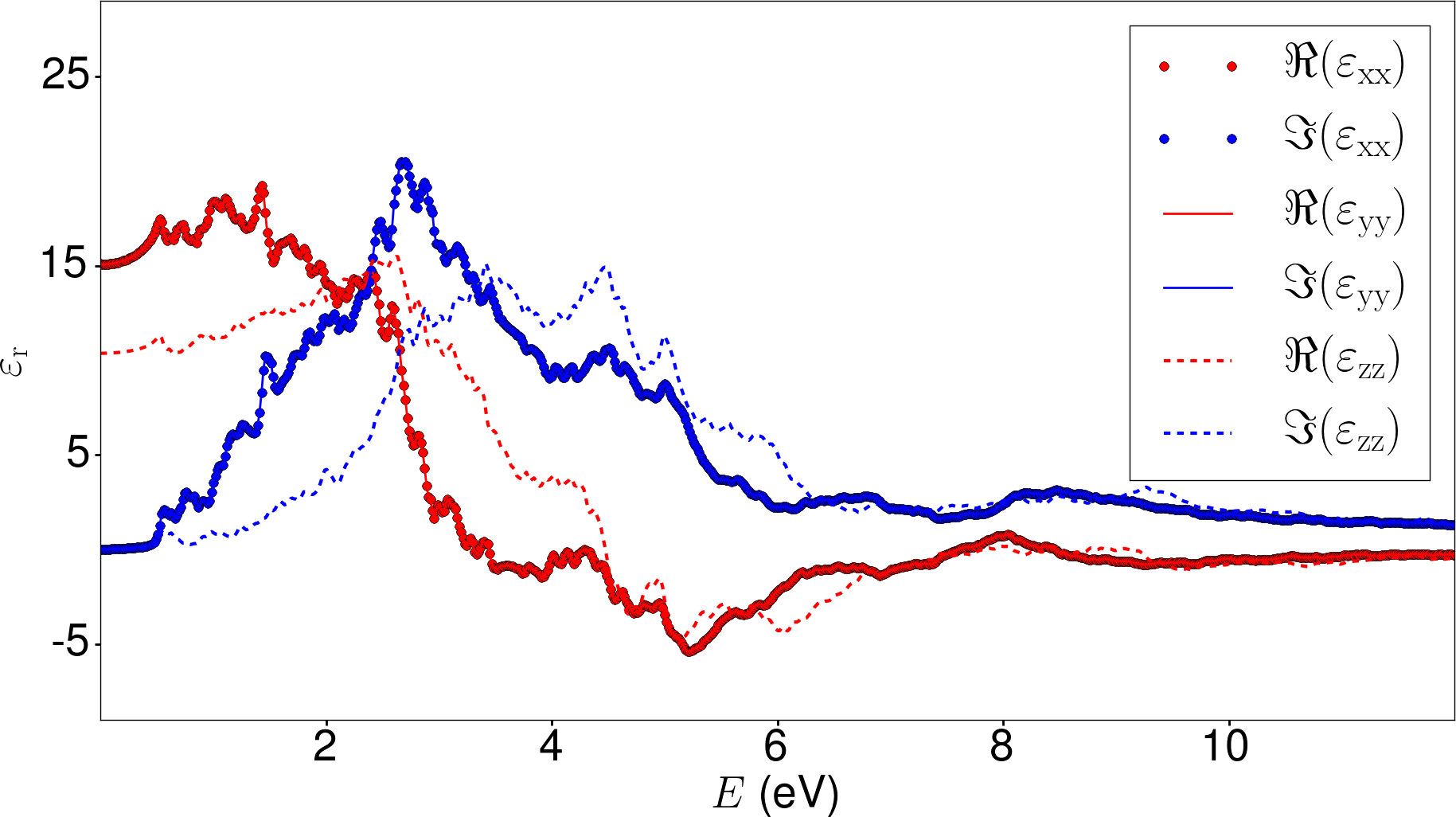}}
 \subfigure[~BiTeBr]{\includegraphics[width=0.329\linewidth]{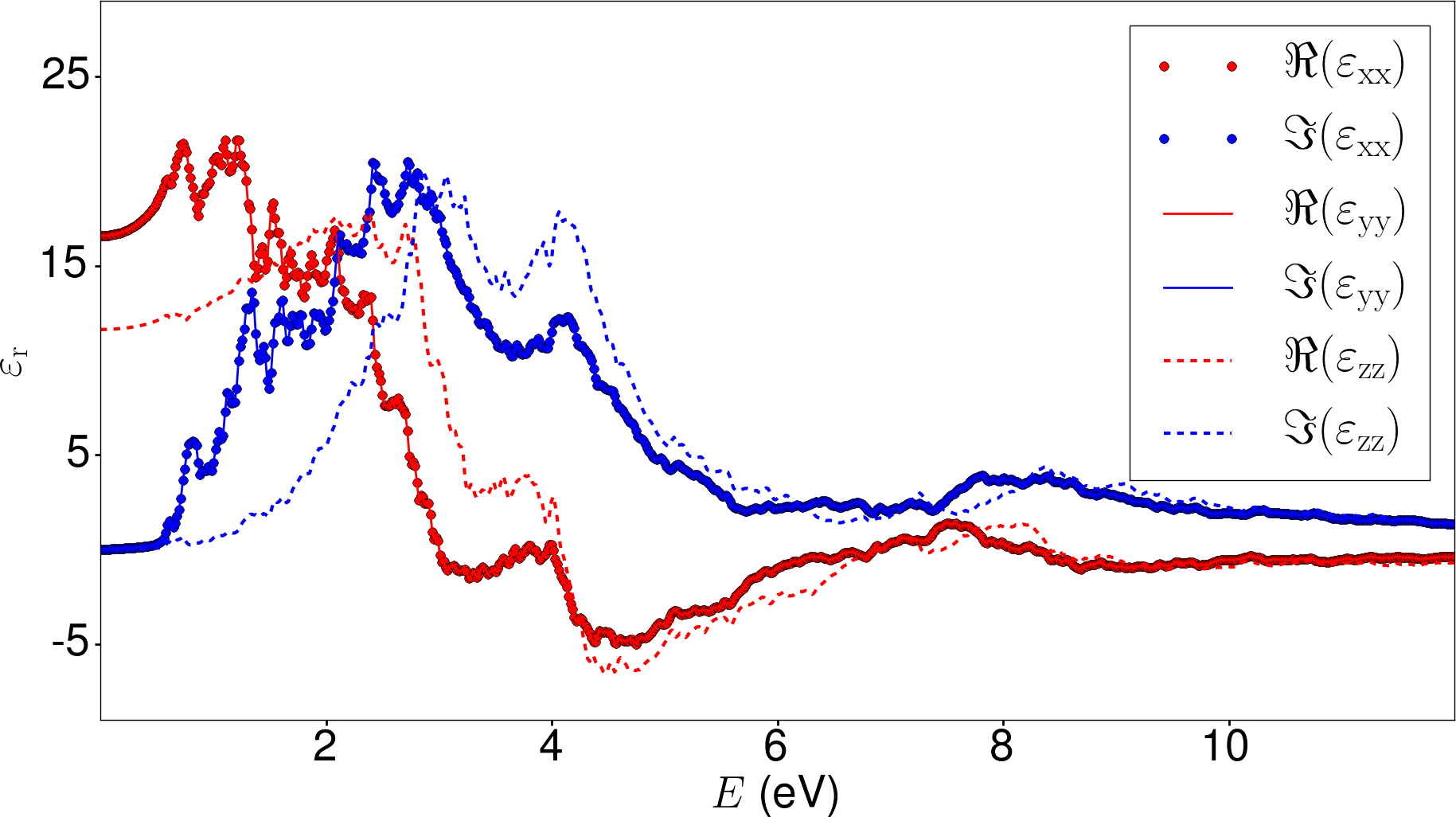}}
 \caption{Dielectric function for BiTeX (X = I,~Cl,~Br). Red lines represent real and blue lines imaginary parts of the dielectric function.}
 \label{BiTeX_EPSI}
\end{figure}


\twocolumngrid
\FloatBarrier


Similarly as in the works of \citet{Lee} and \citet{Demko}, 
we have determined the {\itshape intra}band ($\alpha$,\,$\beta$) transitions (i.e.~transitions within the spin-split conduction bands) and {\itshape inter}band ($\gamma$,\,$\delta$) transitions (i.e.~transitions between valence and conduction bands) in the optical conductivity tensor. The spectral peaks at 0.4 and 0.6~eV are identical for BiTeI, which precisely corresponds to the said interband transitions ($\gamma$,\,$\delta$).\cite{Lee}
The corresponding peaks appear also in BiTeCl and BiTeBr, but their magnitudes and positions differ among these two bismuth tellurohalides. These intra- and interband transitions have also been detected experimentally by \citet{Akrap} (see Fig.~4 there). We remark that the optical transitions between bands with different spin polarization are theoretically expected to occur as a consequence of the SOI.\cite{Lee, Sakano}


\begin{figure}[ht!]
 \centering
 \subfigure[~BiTeI bandstruture]{\includegraphics[width=0.9\linewidth]{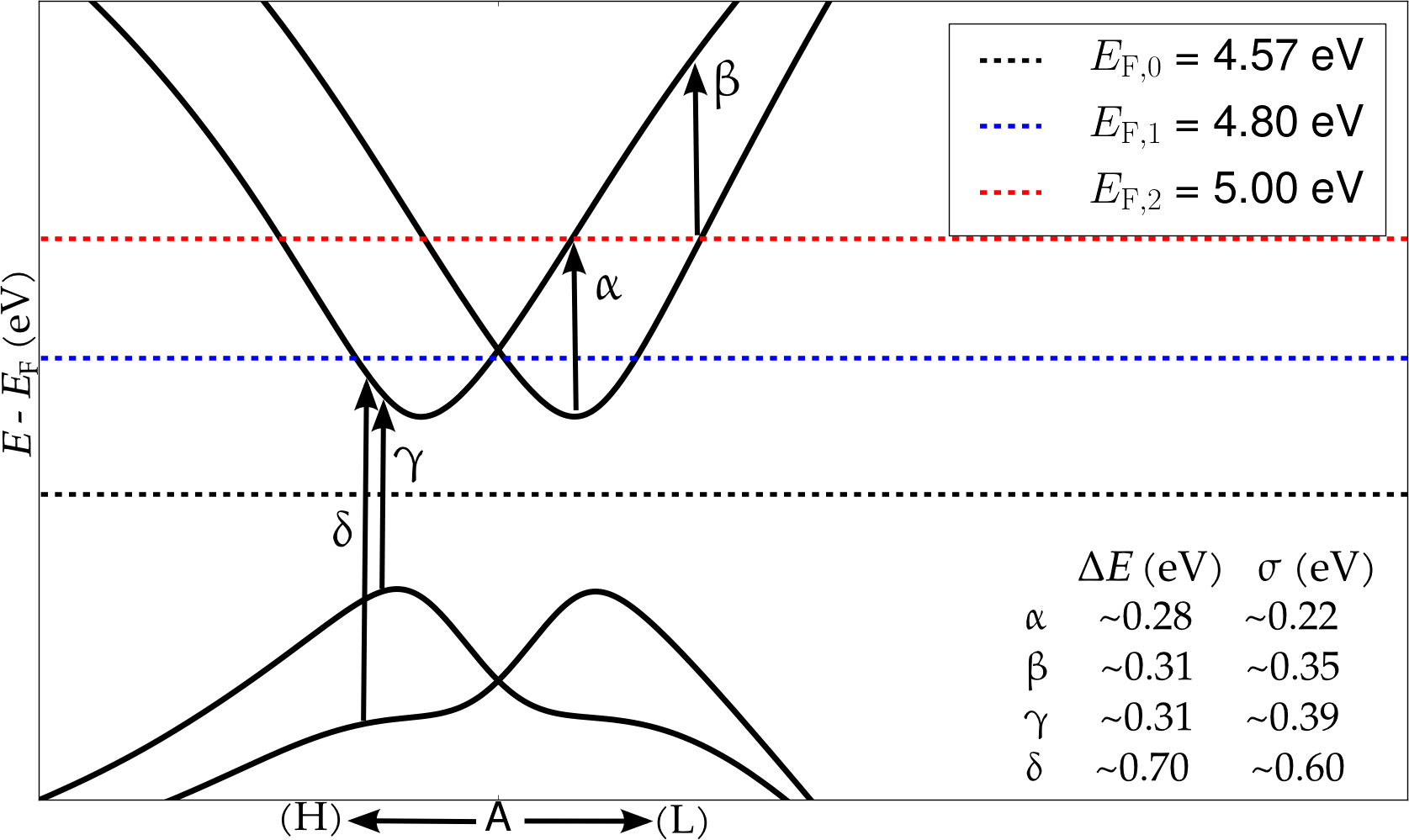}}
 \subfigure[~BiTeI optical conductivity]{\includegraphics[width=0.9\linewidth]{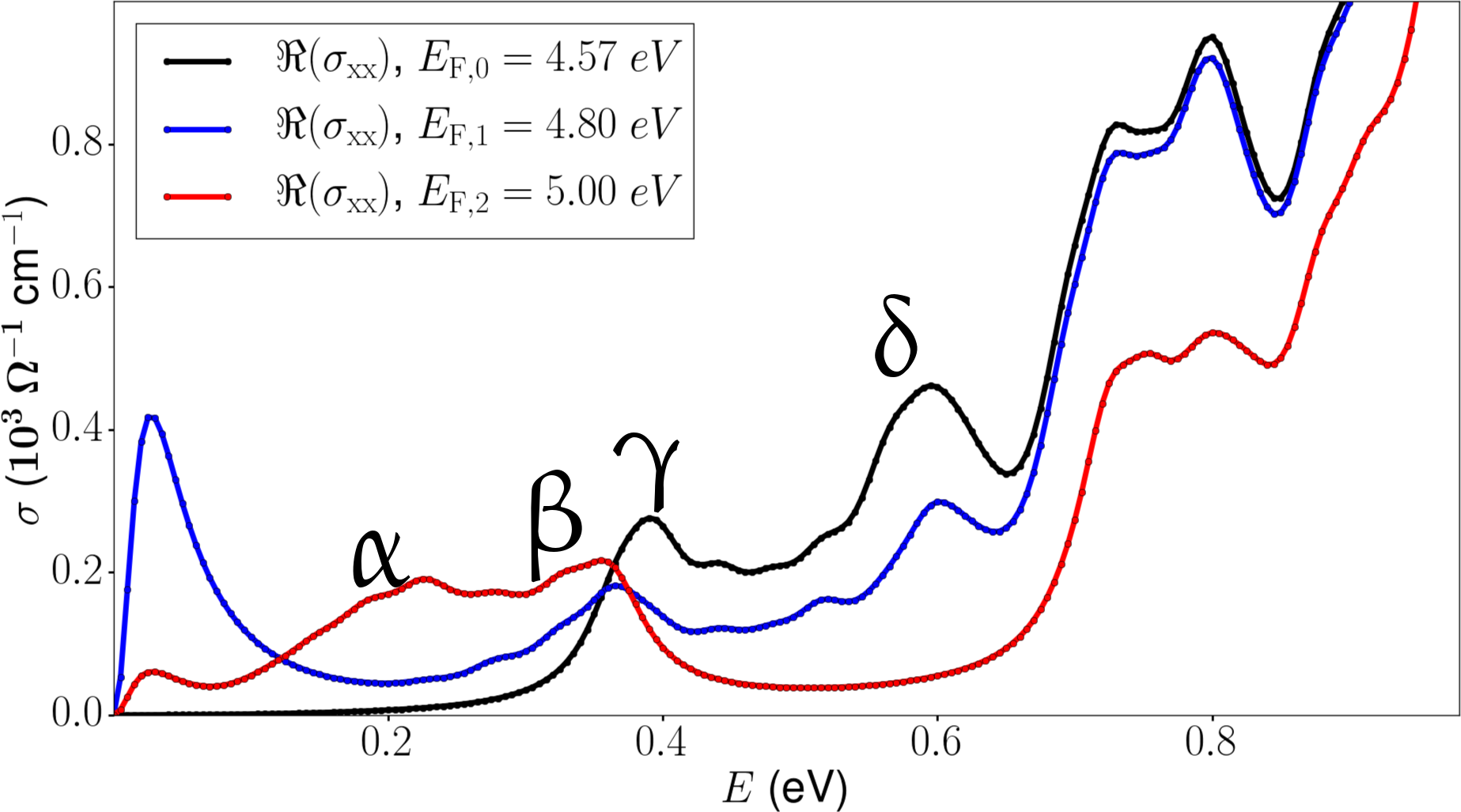}}
 \caption{Real parts of the optical conductivity for BiTeI in dependency of the Fermi level $E_{\text{F}}$. For illustration we have plotted in (a) the different Fermi levels in the band structure of BiTeI, 
 where the same colors (black: $E_{\text{F},0}$ = 4.57 eV, blue: $E_{\text{F},1}$ = 4.80 eV, red: $E_{\text{F},2}$ = 5.00 eV) for the Fermi level are used in (b) for the corresponding optical conductivity results.  In subfigure (b) the characteristic energies correspond to intraband transitions ($\alpha$,\,$\beta$) and interband transitions ($\gamma$,\,$\delta$) as shown schematically in subfigure (a). \label{BiTeI_Ef}}
\end{figure}



Specifically in the case of BiTeI, we have also calculated the optical conductivity (see Fig.~\ref{BiTeI_Ef}(b)) for three different Fermi levels ($E_{\text{F},0}$, $E_{\text{F},1}$, $E_{\text{F},2}$) to simulate the effect of different doping levels. For the highest Fermi level $E_{\text{F},2}$ = 5.0 eV we obtain intraband transitions ($\alpha$, $\beta$) at about 0.2 and 0.35 eV. 
By contrast, in the case of the lowest Fermi level,
the first peak in the optical conductivity (see Fig.~\ref{BiTeI_Ef}(b)) near 0.39 eV corresponds to the electronic band gap of 0.37 eV, 
while the second peak localized near 0.59 eV corresponds to the transition at the A point (see Fig.~\ref{BiTeI_Ef}(a)). 
Generally, by inspection of Table~\ref{tab_E_F}, we observe a rapprochement of the $\gamma$ and $\delta$ peaks with decreasing Fermi level.
Furthermore, we read off that the $\gamma$-peak roughly coincides with the electronic band gap (compare Table~\ref{tab_E_F} with Figs.~\ref{BiTeX_sigma}(a)--(c)).
Note that BiTeCl displays only one peak at 0.57 eV, which is slightly different from the electronic gap whose value is 0.5 eV. 
Fittingly, the transition at the $\Gamma$ point has a value of 0.61 eV.

In addition to the conductivity, we have calculated by means of Eq.~\eqref{eq_use} the dielectric function for all bismuth tellurohalides (see Fig.~\ref{BiTeX_EPSI})
and the resulting dielectric constant (see Table~\ref{tab_n}). Our results for BiTeI and BiTeCl turn out to be in good agreement with the values given in the work of \citet{Rusinov15}.
Correspondingly, we have also calculated the refractive index by means of the relation $n(\omega) = \sqrt{\varepsilon_{\rm r}(\omega)}$ (see Table~\ref{tab_n}) for BiTeX (X = I,~Cl,~Br). For BiTeI and BiTeCl our results are in good agreement with 
the experimental values reported by \citet{Rusinov15}. Moreover, our results for BiTeBr agree with the theoretical values predicted by \citet{BiTeBr_optics}. 


\section{\label{sec:conclusion}Conclusion}

We have shown that the bismuth tellurohalides BiTeI, BiTeCl and BiTeBr, which are highly anisotropic materials with spin-split energy bands, can be reliably treated within DFT (see Fig.~\ref{PDP_FIG}). Moreover, we have shown that all bismuth tellurohalides display a layered electronic localization (see Fig.~\ref{BiTeX_ELF}). Correspondingly, while two of the diagonal elements of the optical conductivity ($\sigma_{xx}$ and $\sigma_{yy}$) display a similar behaviour as a function of the frequency, 
the contribution transverse to the electron localization layer, i.e.~$\sigma_{zz}$, is significantly smaller in the low-frequency region. A further central result of this work is the identification of Rashba-specific transitions (see Fig.~\ref{BiTeX_sigma} and Fig.~\ref{BiTeI_Ef}) within the low frequency branch of the optical conductivity of all three bismuth tellurohalides.

Our calculations of the optical conductivity complement and extend the theoretical results of \citet{Lee} in the following respects: (i) The optical conductivity has been calculated from first principles using all-electron DFT as implemented in the ELK \cite{ELK} code, (ii) we have calculated the whole conductivity tensor (including $\sigma_{zz}$) in both the low-frequency and high-frequency range, and (iii) we have extended the calculation to the compounds BiTeBr and BiTeCl. On the other hand, our results confirm the experimental findings of \citet{Akrap}, where optical transitions were observed even for BiTeBr and BiTeCl, which have a smaller RSS as compared to BiTeI. Furthermore, they confirm that the high-frequency optical spectra of BiTeBr and BiTeCl are similar despite their different space groups.\cite{Akrap} Finally, the optical conductivity of BiTeI as calculated from DFT agrees well with the recent experimental results of \citet{Makhnev} over a wide frequency range. Thus, this work contributes to the understanding of the electron dynamics in the Rashba semiconductors BiTeX (X = I, Cl, Br). Moreover, as optical transitions between Rashba-split bands may be relevant for the resonant dynamical magnetoelectric effect and the spin Hall effect, \cite{Lee} this work also confirms the bismuth tellurohalides as promising candidates for future spintronic applications.

\section*{\label{sec:acknowledgement}Acknowledgements}

The authors thank the group of theoretical physics at the TU Bergakademie Freiberg for seminal discussions.
Furthermore, we thank the ZIH in Dresden for computational time support. S.\,S.~thanks the
SPP 1473 (WeNDeLIB -- Werkstoffe mit neuem Design f\"ur verbesserte Lithium-Ionen-Batterien) for funding. G.\,S.~was supported by the DFG Research Unit FOR723.

\bibliography{rashba,Rashba_Rene,Giulio_Intro}

\begin{thebibliography}{49}%
\makeatletter
\providecommand \@ifxundefined [1]{%
 \@ifx{#1\undefined}
}%
\providecommand \@ifnum [1]{%
 \ifnum #1\expandafter \@firstoftwo
 \else \expandafter \@secondoftwo
 \fi
}%
\providecommand \@ifx [1]{%
 \ifx #1\expandafter \@firstoftwo
 \else \expandafter \@secondoftwo
 \fi
}%
\providecommand \natexlab [1]{#1}%
\providecommand \enquote  [1]{``#1''}%
\providecommand \bibnamefont  [1]{#1}%
\providecommand \bibfnamefont [1]{#1}%
\providecommand \citenamefont [1]{#1}%
\providecommand \href@noop [0]{\@secondoftwo}%
\providecommand \href [0]{\begingroup \@sanitize@url \@href}%
\providecommand \@href[1]{\@@startlink{#1}\@@href}%
\providecommand \@@href[1]{\endgroup#1\@@endlink}%
\providecommand \@sanitize@url [0]{\catcode `\\12\catcode `\$12\catcode
  `\&12\catcode `\#12\catcode `\^12\catcode `\_12\catcode `\%12\relax}%
\providecommand \@@startlink[1]{}%
\providecommand \@@endlink[0]{}%
\providecommand \url  [0]{\begingroup\@sanitize@url \@url }%
\providecommand \@url [1]{\endgroup\@href {#1}{\urlprefix }}%
\providecommand \urlprefix  [0]{URL }%
\providecommand \Eprint [0]{\href }%
\providecommand \doibase [0]{http://dx.doi.org/}%
\providecommand \selectlanguage [0]{\@gobble}%
\providecommand \bibinfo  [0]{\@secondoftwo}%
\providecommand \bibfield  [0]{\@secondoftwo}%
\providecommand \translation [1]{[#1]}%
\providecommand \BibitemOpen [0]{}%
\providecommand \bibitemStop [0]{}%
\providecommand \bibitemNoStop [0]{.\EOS\space}%
\providecommand \EOS [0]{\spacefactor3000\relax}%
\providecommand \BibitemShut  [1]{\csname bibitem#1\endcsname}%
\let\auto@bib@innerbib\@empty
\bibitem [{\citenamefont {Akrap}\ \emph {et~al.}(2014)\citenamefont {Akrap},
  \citenamefont {Teyssier}, \citenamefont {Magrez}, \citenamefont {Bugnon},
  \citenamefont {Berger}, \citenamefont {Kuzmenko},\ and\ \citenamefont
  {van~der Marel}}]{Akrap}%
  \BibitemOpen
  \bibfield  {author} {\bibinfo {author} {\bibfnamefont {A.}~\bibnamefont
  {Akrap}}, \bibinfo {author} {\bibfnamefont {J.}~\bibnamefont {Teyssier}},
  \bibinfo {author} {\bibfnamefont {A.}~\bibnamefont {Magrez}}, \bibinfo
  {author} {\bibfnamefont {P.}~\bibnamefont {Bugnon}}, \bibinfo {author}
  {\bibfnamefont {H.}~\bibnamefont {Berger}}, \bibinfo {author} {\bibfnamefont
  {A.~B.}\ \bibnamefont {Kuzmenko}}, \ and\ \bibinfo {author} {\bibfnamefont
  {D.}~\bibnamefont {van~der Marel}},\ }\href {\doibase
  10.1103/PhysRevB.90.035201} {\bibfield  {journal} {\bibinfo  {journal} {Phys.
  Rev. B}\ }\textbf {\bibinfo {volume} {90}},\ \bibinfo {pages} {035201}
  (\bibinfo {year} {2014})}\BibitemShut {NoStop}%
\bibitem [{\citenamefont {Makhnev}\ \emph {et~al.}(2014)\citenamefont
  {Makhnev}, \citenamefont {Nomerovannaya}, \citenamefont {Kuznetsova},
  \citenamefont {Tereshchenko},\ and\ \citenamefont {Kokh}}]{Makhnev}%
  \BibitemOpen
  \bibfield  {author} {\bibinfo {author} {\bibfnamefont {A.~A.}\ \bibnamefont
  {Makhnev}}, \bibinfo {author} {\bibfnamefont {L.~V.}\ \bibnamefont
  {Nomerovannaya}}, \bibinfo {author} {\bibfnamefont {T.~V.}\ \bibnamefont
  {Kuznetsova}}, \bibinfo {author} {\bibfnamefont {O.~E.}\ \bibnamefont
  {Tereshchenko}}, \ and\ \bibinfo {author} {\bibfnamefont {K.~A.}\
  \bibnamefont {Kokh}},\ }\href@noop {} {\bibfield  {journal} {\bibinfo
  {journal} {Optics and Spectroscopy}\ }\textbf {\bibinfo {volume} {117}},\
  \bibinfo {pages} {764} (\bibinfo {year} {2014})},\ \bibinfo {note} {[Optika i
  Spektroskopiya {\bfseries 117}, 789--793 (2014)]}\BibitemShut {NoStop}%
\bibitem [{\citenamefont {Rusinov}\ \emph {et~al.}(2015)\citenamefont
  {Rusinov}, \citenamefont {Tereshchenko}, \citenamefont {Kokh}, \citenamefont
  {Shakhmametova}, \citenamefont {Azarov},\ and\ \citenamefont
  {Chulkov}}]{Rusinov15}%
  \BibitemOpen
  \bibfield  {author} {\bibinfo {author} {\bibfnamefont {I.~P.}\ \bibnamefont
  {Rusinov}}, \bibinfo {author} {\bibfnamefont {O.~E.}\ \bibnamefont
  {Tereshchenko}}, \bibinfo {author} {\bibfnamefont {K.~A.}\ \bibnamefont
  {Kokh}}, \bibinfo {author} {\bibfnamefont {A.~R.}\ \bibnamefont
  {Shakhmametova}}, \bibinfo {author} {\bibfnamefont {I.~A.}\ \bibnamefont
  {Azarov}}, \ and\ \bibinfo {author} {\bibfnamefont {E.~V.}\ \bibnamefont
  {Chulkov}},\ }\href@noop {} {\bibfield  {journal} {\bibinfo  {journal} {JETP
  Letters}\ }\textbf {\bibinfo {volume} {101}},\ \bibinfo {pages} {507}
  (\bibinfo {year} {2015})},\ \bibinfo {note} {[Pis'ma v Zhurnal
  Eksperimental'noi i Teoreticheskoi Fiziki {\bfseries 101}, 563--568
  (2015)]}\BibitemShut {NoStop}%
\bibitem [{\citenamefont {Sinova}\ and\ \citenamefont
  {{\v{Z}uti\'{c}}}(2012)}]{Sinova}%
  \BibitemOpen
  \bibfield  {author} {\bibinfo {author} {\bibfnamefont {J.}~\bibnamefont
  {Sinova}}\ and\ \bibinfo {author} {\bibfnamefont {I.}~\bibnamefont
  {{\v{Z}uti\'{c}}}},\ }\href {http://dx.doi.org/10.1038/nmat3304} {\bibfield
  {journal} {\bibinfo  {journal} {Nat. Mater.}\ }\textbf {\bibinfo {volume}
  {11}},\ \bibinfo {pages} {368} (\bibinfo {year} {2012})}\BibitemShut
  {NoStop}%
\bibitem [{\citenamefont {Barnes}\ \emph {et~al.}(2014)\citenamefont {Barnes},
  \citenamefont {Ieda},\ and\ \citenamefont {Maekawa}}]{Barnes}%
  \BibitemOpen
  \bibfield  {author} {\bibinfo {author} {\bibfnamefont {S.~E.}\ \bibnamefont
  {Barnes}}, \bibinfo {author} {\bibfnamefont {J.}~\bibnamefont {Ieda}}, \ and\
  \bibinfo {author} {\bibfnamefont {S.}~\bibnamefont {Maekawa}},\ }\href
  {http://dx.doi.org/10.1038/srep04105} {\bibfield  {journal} {\bibinfo
  {journal} {Sci. Rep.}\ }\textbf {\bibinfo {volume} {4}},\ \bibinfo {pages}
  {4105} (\bibinfo {year} {2014})}\BibitemShut {NoStop}%
\bibitem [{\citenamefont {Rashba}(1960)}]{Rashba60}%
  \BibitemOpen
  \bibfield  {author} {\bibinfo {author} {\bibfnamefont {E.~I.}\ \bibnamefont
  {Rashba}},\ }\href@noop {} {\bibfield  {journal} {\bibinfo  {journal} {Sov.
  Phys.--Solid State}\ }\textbf {\bibinfo {volume} {2}},\ \bibinfo {pages}
  {1109} (\bibinfo {year} {1960})},\ \bibinfo {note} {[Fizika Tverd. Tela
  {\bfseries 2}, 1224 (1960)]}\BibitemShut {NoStop}%
\bibitem [{\citenamefont {LaShell}\ \emph {et~al.}(1996)\citenamefont
  {LaShell}, \citenamefont {McDougall},\ and\ \citenamefont
  {Jensen}}]{LaShell96}%
  \BibitemOpen
  \bibfield  {author} {\bibinfo {author} {\bibfnamefont {S.}~\bibnamefont
  {LaShell}}, \bibinfo {author} {\bibfnamefont {B.~A.}\ \bibnamefont
  {McDougall}}, \ and\ \bibinfo {author} {\bibfnamefont {E.}~\bibnamefont
  {Jensen}},\ }\href {\doibase 10.1103/PhysRevLett.77.3419} {\bibfield
  {journal} {\bibinfo  {journal} {Phys. Rev. Lett.}\ }\textbf {\bibinfo
  {volume} {77}},\ \bibinfo {pages} {3419} (\bibinfo {year}
  {1996})}\BibitemShut {NoStop}%
\bibitem [{\citenamefont {Nitta}\ \emph {et~al.}(1997)\citenamefont {Nitta},
  \citenamefont {Akazaki}, \citenamefont {Takayanagi},\ and\ \citenamefont
  {Enoki}}]{Nitta97}%
  \BibitemOpen
  \bibfield  {author} {\bibinfo {author} {\bibfnamefont {J.}~\bibnamefont
  {Nitta}}, \bibinfo {author} {\bibfnamefont {T.}~\bibnamefont {Akazaki}},
  \bibinfo {author} {\bibfnamefont {H.}~\bibnamefont {Takayanagi}}, \ and\
  \bibinfo {author} {\bibfnamefont {T.}~\bibnamefont {Enoki}},\ }\href
  {\doibase 10.1103/PhysRevLett.78.1335} {\bibfield  {journal} {\bibinfo
  {journal} {Phys. Rev. Lett.}\ }\textbf {\bibinfo {volume} {78}},\ \bibinfo
  {pages} {1335} (\bibinfo {year} {1997})}\BibitemShut {NoStop}%
\bibitem [{\citenamefont {Ast}\ \emph {et~al.}(2007)\citenamefont {Ast},
  \citenamefont {Henk}, \citenamefont {Ernst}, \citenamefont {Moreschini},
  \citenamefont {Falub}, \citenamefont {Pacil\'e}, \citenamefont {Bruno},
  \citenamefont {Kern},\ and\ \citenamefont {Grioni}}]{Ast07}%
  \BibitemOpen
  \bibfield  {author} {\bibinfo {author} {\bibfnamefont {C.~R.}\ \bibnamefont
  {Ast}}, \bibinfo {author} {\bibfnamefont {J.}~\bibnamefont {Henk}}, \bibinfo
  {author} {\bibfnamefont {A.}~\bibnamefont {Ernst}}, \bibinfo {author}
  {\bibfnamefont {L.}~\bibnamefont {Moreschini}}, \bibinfo {author}
  {\bibfnamefont {M.~C.}\ \bibnamefont {Falub}}, \bibinfo {author}
  {\bibfnamefont {D.}~\bibnamefont {Pacil\'e}}, \bibinfo {author}
  {\bibfnamefont {P.}~\bibnamefont {Bruno}}, \bibinfo {author} {\bibfnamefont
  {K.}~\bibnamefont {Kern}}, \ and\ \bibinfo {author} {\bibfnamefont
  {M.}~\bibnamefont {Grioni}},\ }\href {\doibase 10.1103/PhysRevLett.98.186807}
  {\bibfield  {journal} {\bibinfo  {journal} {Phys. Rev. Lett.}\ }\textbf
  {\bibinfo {volume} {98}},\ \bibinfo {pages} {186807} (\bibinfo {year}
  {2007})}\BibitemShut {NoStop}%
\bibitem [{\citenamefont {Marchenko}\ \emph {et~al.}(2012)\citenamefont
  {Marchenko}, \citenamefont {Varykhalov}, \citenamefont {Scholz},
  \citenamefont {Bihlmayer}, \citenamefont {Rashba}, \citenamefont {Rybkin},
  \citenamefont {Shikin},\ and\ \citenamefont {Rader}}]{2012giant}%
  \BibitemOpen
  \bibfield  {author} {\bibinfo {author} {\bibfnamefont {D.}~\bibnamefont
  {Marchenko}}, \bibinfo {author} {\bibfnamefont {A.}~\bibnamefont
  {Varykhalov}}, \bibinfo {author} {\bibfnamefont {M.}~\bibnamefont {Scholz}},
  \bibinfo {author} {\bibfnamefont {G.}~\bibnamefont {Bihlmayer}}, \bibinfo
  {author} {\bibfnamefont {E.}~\bibnamefont {Rashba}}, \bibinfo {author}
  {\bibfnamefont {A.}~\bibnamefont {Rybkin}}, \bibinfo {author} {\bibfnamefont
  {A.}~\bibnamefont {Shikin}}, \ and\ \bibinfo {author} {\bibfnamefont
  {O.}~\bibnamefont {Rader}},\ }\href@noop {} {\bibfield  {journal} {\bibinfo
  {journal} {Nat. Commun.}\ }\textbf {\bibinfo {volume} {3}},\ \bibinfo {pages}
  {1232} (\bibinfo {year} {2012})}\BibitemShut {NoStop}%
\bibitem [{\citenamefont {Ishizaka}\ \emph {et~al.}(2011)\citenamefont
  {Ishizaka}, \citenamefont {Bahramy}, \citenamefont {Murakawa}, \citenamefont
  {Sakano}, \citenamefont {Shimojima}, \citenamefont {Sonobe}, \citenamefont
  {Koizumi}, \citenamefont {Shin}, \citenamefont {Miyahara}, \citenamefont
  {Kimura} \emph {et~al.}}]{2011giant}%
  \BibitemOpen
  \bibfield  {author} {\bibinfo {author} {\bibfnamefont {K.}~\bibnamefont
  {Ishizaka}}, \bibinfo {author} {\bibfnamefont {M.}~\bibnamefont {Bahramy}},
  \bibinfo {author} {\bibfnamefont {H.}~\bibnamefont {Murakawa}}, \bibinfo
  {author} {\bibfnamefont {M.}~\bibnamefont {Sakano}}, \bibinfo {author}
  {\bibfnamefont {T.}~\bibnamefont {Shimojima}}, \bibinfo {author}
  {\bibfnamefont {T.}~\bibnamefont {Sonobe}}, \bibinfo {author} {\bibfnamefont
  {K.}~\bibnamefont {Koizumi}}, \bibinfo {author} {\bibfnamefont
  {S.}~\bibnamefont {Shin}}, \bibinfo {author} {\bibfnamefont {H.}~\bibnamefont
  {Miyahara}}, \bibinfo {author} {\bibfnamefont {A.}~\bibnamefont {Kimura}},
  \emph {et~al.},\ }\href@noop {} {\bibfield  {journal} {\bibinfo  {journal}
  {Nat. Mater.}\ }\textbf {\bibinfo {volume} {10}},\ \bibinfo {pages} {521}
  (\bibinfo {year} {2011})}\BibitemShut {NoStop}%
\bibitem [{\citenamefont {Crepaldi}\ \emph {et~al.}(2012)\citenamefont
  {Crepaldi}, \citenamefont {Moreschini}, \citenamefont {Aut\`es},
  \citenamefont {Tournier-Colletta}, \citenamefont {Moser}, \citenamefont
  {Virk}, \citenamefont {Berger}, \citenamefont {Bugnon}, \citenamefont
  {Chang}, \citenamefont {Kern}, \citenamefont {Bostwick}, \citenamefont
  {Rotenberg}, \citenamefont {Yazyev},\ and\ \citenamefont
  {Grioni}}]{Crepaldi}%
  \BibitemOpen
  \bibfield  {author} {\bibinfo {author} {\bibfnamefont {A.}~\bibnamefont
  {Crepaldi}}, \bibinfo {author} {\bibfnamefont {L.}~\bibnamefont
  {Moreschini}}, \bibinfo {author} {\bibfnamefont {G.}~\bibnamefont {Aut\`es}},
  \bibinfo {author} {\bibfnamefont {C.}~\bibnamefont {Tournier-Colletta}},
  \bibinfo {author} {\bibfnamefont {S.}~\bibnamefont {Moser}}, \bibinfo
  {author} {\bibfnamefont {N.}~\bibnamefont {Virk}}, \bibinfo {author}
  {\bibfnamefont {H.}~\bibnamefont {Berger}}, \bibinfo {author} {\bibfnamefont
  {P.}~\bibnamefont {Bugnon}}, \bibinfo {author} {\bibfnamefont {Y.~J.}\
  \bibnamefont {Chang}}, \bibinfo {author} {\bibfnamefont {K.}~\bibnamefont
  {Kern}}, \bibinfo {author} {\bibfnamefont {A.}~\bibnamefont {Bostwick}},
  \bibinfo {author} {\bibfnamefont {E.}~\bibnamefont {Rotenberg}}, \bibinfo
  {author} {\bibfnamefont {O.~V.}\ \bibnamefont {Yazyev}}, \ and\ \bibinfo
  {author} {\bibfnamefont {M.}~\bibnamefont {Grioni}},\ }\href {\doibase
  10.1103/PhysRevLett.109.096803} {\bibfield  {journal} {\bibinfo  {journal}
  {Phys. Rev. Lett.}\ }\textbf {\bibinfo {volume} {109}},\ \bibinfo {pages}
  {096803} (\bibinfo {year} {2012})}\BibitemShut {NoStop}%
\bibitem [{\citenamefont {Lee}\ \emph {et~al.}(2011)\citenamefont {Lee},
  \citenamefont {Schober}, \citenamefont {Bahramy}, \citenamefont {Murakawa},
  \citenamefont {Onose}, \citenamefont {Arita}, \citenamefont {Nagaosa},\ and\
  \citenamefont {Tokura}}]{Lee}%
  \BibitemOpen
  \bibfield  {author} {\bibinfo {author} {\bibfnamefont {J.~S.}\ \bibnamefont
  {Lee}}, \bibinfo {author} {\bibfnamefont {G.~A.~H.}\ \bibnamefont {Schober}},
  \bibinfo {author} {\bibfnamefont {M.~S.}\ \bibnamefont {Bahramy}}, \bibinfo
  {author} {\bibfnamefont {H.}~\bibnamefont {Murakawa}}, \bibinfo {author}
  {\bibfnamefont {Y.}~\bibnamefont {Onose}}, \bibinfo {author} {\bibfnamefont
  {R.}~\bibnamefont {Arita}}, \bibinfo {author} {\bibfnamefont
  {N.}~\bibnamefont {Nagaosa}}, \ and\ \bibinfo {author} {\bibfnamefont
  {Y.}~\bibnamefont {Tokura}},\ }\href {\doibase
  10.1103/PhysRevLett.107.117401} {\bibfield  {journal} {\bibinfo  {journal}
  {Phys. Rev. Lett.}\ }\textbf {\bibinfo {volume} {107}},\ \bibinfo {pages}
  {117401} (\bibinfo {year} {2011})}\BibitemShut {NoStop}%
\bibitem [{\citenamefont {Schober}\ \emph {et~al.}(2012)\citenamefont
  {Schober}, \citenamefont {Murakawa}, \citenamefont {Bahramy}, \citenamefont
  {Arita}, \citenamefont {Kaneko}, \citenamefont {Tokura},\ and\ \citenamefont
  {Nagaosa}}]{Schober}%
  \BibitemOpen
  \bibfield  {author} {\bibinfo {author} {\bibfnamefont {G.~A.~H.}\
  \bibnamefont {Schober}}, \bibinfo {author} {\bibfnamefont {H.}~\bibnamefont
  {Murakawa}}, \bibinfo {author} {\bibfnamefont {M.~S.}\ \bibnamefont
  {Bahramy}}, \bibinfo {author} {\bibfnamefont {R.}~\bibnamefont {Arita}},
  \bibinfo {author} {\bibfnamefont {Y.}~\bibnamefont {Kaneko}}, \bibinfo
  {author} {\bibfnamefont {Y.}~\bibnamefont {Tokura}}, \ and\ \bibinfo {author}
  {\bibfnamefont {N.}~\bibnamefont {Nagaosa}},\ }\href {\doibase
  10.1103/PhysRevLett.108.247208} {\bibfield  {journal} {\bibinfo  {journal}
  {Phys. Rev. Lett.}\ }\textbf {\bibinfo {volume} {108}},\ \bibinfo {pages}
  {247208} (\bibinfo {year} {2012})}\BibitemShut {NoStop}%
\bibitem [{\citenamefont {Demk\'o}\ \emph {et~al.}(2012)\citenamefont
  {Demk\'o}, \citenamefont {Schober}, \citenamefont {Kocsis}, \citenamefont
  {Bahramy}, \citenamefont {Murakawa}, \citenamefont {Lee}, \citenamefont
  {K\'ezsm\'arki}, \citenamefont {Arita}, \citenamefont {Nagaosa},\ and\
  \citenamefont {Tokura}}]{Demko}%
  \BibitemOpen
  \bibfield  {author} {\bibinfo {author} {\bibfnamefont {L.}~\bibnamefont
  {Demk\'o}}, \bibinfo {author} {\bibfnamefont {G.~A.~H.}\ \bibnamefont
  {Schober}}, \bibinfo {author} {\bibfnamefont {V.}~\bibnamefont {Kocsis}},
  \bibinfo {author} {\bibfnamefont {M.~S.}\ \bibnamefont {Bahramy}}, \bibinfo
  {author} {\bibfnamefont {H.}~\bibnamefont {Murakawa}}, \bibinfo {author}
  {\bibfnamefont {J.~S.}\ \bibnamefont {Lee}}, \bibinfo {author} {\bibfnamefont
  {I.}~\bibnamefont {K\'ezsm\'arki}}, \bibinfo {author} {\bibfnamefont
  {R.}~\bibnamefont {Arita}}, \bibinfo {author} {\bibfnamefont
  {N.}~\bibnamefont {Nagaosa}}, \ and\ \bibinfo {author} {\bibfnamefont
  {Y.}~\bibnamefont {Tokura}},\ }\href {\doibase
  10.1103/PhysRevLett.109.167401} {\bibfield  {journal} {\bibinfo  {journal}
  {Phys. Rev. Lett.}\ }\textbf {\bibinfo {volume} {109}},\ \bibinfo {pages}
  {167401} (\bibinfo {year} {2012})}\BibitemShut {NoStop}%
\bibitem [{\citenamefont {Bahramy}\ \emph {et~al.}(2012)\citenamefont
  {Bahramy}, \citenamefont {Yang}, \citenamefont {Arita},\ and\ \citenamefont
  {Nagaosa}}]{BiTeITop}%
  \BibitemOpen
  \bibfield  {author} {\bibinfo {author} {\bibfnamefont {M.~S.}\ \bibnamefont
  {Bahramy}}, \bibinfo {author} {\bibfnamefont {B.-J.}\ \bibnamefont {Yang}},
  \bibinfo {author} {\bibfnamefont {R.}~\bibnamefont {Arita}}, \ and\ \bibinfo
  {author} {\bibfnamefont {N.}~\bibnamefont {Nagaosa}},\ }\href@noop {}
  {\bibfield  {journal} {\bibinfo  {journal} {Nat. Commun.}\ }\textbf {\bibinfo
  {volume} {3}},\ \bibinfo {pages} {679} (\bibinfo {year} {2012})}\BibitemShut
  {NoStop}%
\bibitem [{\citenamefont {Xi}\ \emph {et~al.}(2013)\citenamefont {Xi},
  \citenamefont {Ma}, \citenamefont {Liu}, \citenamefont {Chen}, \citenamefont
  {Ku}, \citenamefont {Berger}, \citenamefont {Martin}, \citenamefont
  {Tanner},\ and\ \citenamefont {Carr}}]{Xi13}%
  \BibitemOpen
  \bibfield  {author} {\bibinfo {author} {\bibfnamefont {X.}~\bibnamefont
  {Xi}}, \bibinfo {author} {\bibfnamefont {C.}~\bibnamefont {Ma}}, \bibinfo
  {author} {\bibfnamefont {Z.}~\bibnamefont {Liu}}, \bibinfo {author}
  {\bibfnamefont {Z.}~\bibnamefont {Chen}}, \bibinfo {author} {\bibfnamefont
  {W.}~\bibnamefont {Ku}}, \bibinfo {author} {\bibfnamefont {H.}~\bibnamefont
  {Berger}}, \bibinfo {author} {\bibfnamefont {C.}~\bibnamefont {Martin}},
  \bibinfo {author} {\bibfnamefont {D.~B.}\ \bibnamefont {Tanner}}, \ and\
  \bibinfo {author} {\bibfnamefont {G.~L.}\ \bibnamefont {Carr}},\ }\href
  {\doibase 10.1103/PhysRevLett.111.155701} {\bibfield  {journal} {\bibinfo
  {journal} {Phys. Rev. Lett.}\ }\textbf {\bibinfo {volume} {111}},\ \bibinfo
  {pages} {155701} (\bibinfo {year} {2013})}\BibitemShut {NoStop}%
\bibitem [{\citenamefont {Tran}\ \emph {et~al.}(2014)\citenamefont {Tran},
  \citenamefont {Levallois}, \citenamefont {Lerch}, \citenamefont {Teyssier},
  \citenamefont {Kuzmenko}, \citenamefont {Aut\`es}, \citenamefont {Yazyev},
  \citenamefont {Ubaldini}, \citenamefont {Giannini}, \citenamefont {van~der
  Marel},\ and\ \citenamefont {Akrap}}]{Tran14}%
  \BibitemOpen
  \bibfield  {author} {\bibinfo {author} {\bibfnamefont {M.~K.}\ \bibnamefont
  {Tran}}, \bibinfo {author} {\bibfnamefont {J.}~\bibnamefont {Levallois}},
  \bibinfo {author} {\bibfnamefont {P.}~\bibnamefont {Lerch}}, \bibinfo
  {author} {\bibfnamefont {J.}~\bibnamefont {Teyssier}}, \bibinfo {author}
  {\bibfnamefont {A.~B.}\ \bibnamefont {Kuzmenko}}, \bibinfo {author}
  {\bibfnamefont {G.}~\bibnamefont {Aut\`es}}, \bibinfo {author} {\bibfnamefont
  {O.~V.}\ \bibnamefont {Yazyev}}, \bibinfo {author} {\bibfnamefont
  {A.}~\bibnamefont {Ubaldini}}, \bibinfo {author} {\bibfnamefont
  {E.}~\bibnamefont {Giannini}}, \bibinfo {author} {\bibfnamefont
  {D.}~\bibnamefont {van~der Marel}}, \ and\ \bibinfo {author} {\bibfnamefont
  {A.}~\bibnamefont {Akrap}},\ }\href {\doibase 10.1103/PhysRevLett.112.047402}
  {\bibfield  {journal} {\bibinfo  {journal} {Phys. Rev. Lett.}\ }\textbf
  {\bibinfo {volume} {112}},\ \bibinfo {pages} {047402} (\bibinfo {year}
  {2014})}\BibitemShut {NoStop}%
\bibitem [{\citenamefont {Park}\ \emph {et~al.}(2015)\citenamefont {Park},
  \citenamefont {Jin}, \citenamefont {Jo}, \citenamefont {Choi}, \citenamefont
  {Kang}, \citenamefont {Kampert}, \citenamefont {Rhyee}, \citenamefont {Jhi},\
  and\ \citenamefont {Kim}}]{Park15}%
  \BibitemOpen
  \bibfield  {author} {\bibinfo {author} {\bibfnamefont {J.}~\bibnamefont
  {Park}}, \bibinfo {author} {\bibfnamefont {K.-H.}\ \bibnamefont {Jin}},
  \bibinfo {author} {\bibfnamefont {Y.~J.}\ \bibnamefont {Jo}}, \bibinfo
  {author} {\bibfnamefont {E.~S.}\ \bibnamefont {Choi}}, \bibinfo {author}
  {\bibfnamefont {W.}~\bibnamefont {Kang}}, \bibinfo {author} {\bibfnamefont
  {E.}~\bibnamefont {Kampert}}, \bibinfo {author} {\bibfnamefont {J.-S.}\
  \bibnamefont {Rhyee}}, \bibinfo {author} {\bibfnamefont {S.-H.}\ \bibnamefont
  {Jhi}}, \ and\ \bibinfo {author} {\bibfnamefont {J.~S.}\ \bibnamefont
  {Kim}},\ }\href@noop {} {\bibfield  {journal} {\bibinfo  {journal} {Sci.
  Rep.}\ }\textbf {\bibinfo {volume} {5}},\ \bibinfo {pages} {15973} (\bibinfo
  {year} {2015})}\BibitemShut {NoStop}%
\bibitem [{\citenamefont {Bahramy}\ \emph {et~al.}()\citenamefont {Bahramy},
  \citenamefont {Arita},\ and\ \citenamefont {Nagaosa}}]{BiTeI_WIEN2k}%
  \BibitemOpen
  \bibfield  {author} {\bibinfo {author} {\bibfnamefont {M.~S.}\ \bibnamefont
  {Bahramy}}, \bibinfo {author} {\bibfnamefont {R.}~\bibnamefont {Arita}}, \
  and\ \bibinfo {author} {\bibfnamefont {N.}~\bibnamefont {Nagaosa}},\ }\href
  {\doibase 10.1103/PhysRevB.84.041202} {\ \textbf {\bibinfo {volume} {84}},\
  \bibinfo {pages} {041202}}\BibitemShut {NoStop}%
\bibitem [{\citenamefont {Eremeev}\ \emph {et~al.}(2012)\citenamefont
  {Eremeev}, \citenamefont {Nechaev}, \citenamefont {Koroteev}, \citenamefont
  {Echenique},\ and\ \citenamefont {Chulkov}}]{Eremeev}%
  \BibitemOpen
  \bibfield  {author} {\bibinfo {author} {\bibfnamefont {S.~V.}\ \bibnamefont
  {Eremeev}}, \bibinfo {author} {\bibfnamefont {I.~A.}\ \bibnamefont
  {Nechaev}}, \bibinfo {author} {\bibfnamefont {Y.~M.}\ \bibnamefont
  {Koroteev}}, \bibinfo {author} {\bibfnamefont {P.~M.}\ \bibnamefont
  {Echenique}}, \ and\ \bibinfo {author} {\bibfnamefont {E.~V.}\ \bibnamefont
  {Chulkov}},\ }\href {\doibase 10.1103/PhysRevLett.108.246802} {\bibfield
  {journal} {\bibinfo  {journal} {Phys. Rev. Lett.}\ }\textbf {\bibinfo
  {volume} {108}},\ \bibinfo {pages} {246802} (\bibinfo {year}
  {2012})}\BibitemShut {NoStop}%
\bibitem [{\citenamefont {Sakano}\ \emph {et~al.}(2013)\citenamefont {Sakano},
  \citenamefont {Bahramy}, \citenamefont {Katayama}, \citenamefont {Shimojima},
  \citenamefont {Murakawa}, \citenamefont {Kaneko}, \citenamefont {Malaeb},
  \citenamefont {Shin}, \citenamefont {Ono}, \citenamefont {Kumigashira},
  \citenamefont {Arita}, \citenamefont {Nagaosa}, \citenamefont {Hwang},
  \citenamefont {Tokura},\ and\ \citenamefont {Ishizaka}}]{Sakano}%
  \BibitemOpen
  \bibfield  {author} {\bibinfo {author} {\bibfnamefont {M.}~\bibnamefont
  {Sakano}}, \bibinfo {author} {\bibfnamefont {M.~S.}\ \bibnamefont {Bahramy}},
  \bibinfo {author} {\bibfnamefont {A.}~\bibnamefont {Katayama}}, \bibinfo
  {author} {\bibfnamefont {T.}~\bibnamefont {Shimojima}}, \bibinfo {author}
  {\bibfnamefont {H.}~\bibnamefont {Murakawa}}, \bibinfo {author}
  {\bibfnamefont {Y.}~\bibnamefont {Kaneko}}, \bibinfo {author} {\bibfnamefont
  {W.}~\bibnamefont {Malaeb}}, \bibinfo {author} {\bibfnamefont
  {S.}~\bibnamefont {Shin}}, \bibinfo {author} {\bibfnamefont {K.}~\bibnamefont
  {Ono}}, \bibinfo {author} {\bibfnamefont {H.}~\bibnamefont {Kumigashira}},
  \bibinfo {author} {\bibfnamefont {R.}~\bibnamefont {Arita}}, \bibinfo
  {author} {\bibfnamefont {N.}~\bibnamefont {Nagaosa}}, \bibinfo {author}
  {\bibfnamefont {H.~Y.}\ \bibnamefont {Hwang}}, \bibinfo {author}
  {\bibfnamefont {Y.}~\bibnamefont {Tokura}}, \ and\ \bibinfo {author}
  {\bibfnamefont {K.}~\bibnamefont {Ishizaka}},\ }\href {\doibase
  10.1103/PhysRevLett.110.107204} {\bibfield  {journal} {\bibinfo  {journal}
  {Phys. Rev. Lett.}\ }\textbf {\bibinfo {volume} {110}},\ \bibinfo {pages}
  {107204} (\bibinfo {year} {2013})}\BibitemShut {NoStop}%
\bibitem [{\citenamefont {Chen}\ \emph
  {et~al.}(2013{\natexlab{a}})\citenamefont {Chen}, \citenamefont {Kanou},
  \citenamefont {Liu}, \citenamefont {Zhang}, \citenamefont {Sobota},
  \citenamefont {Leuenberger}, \citenamefont {Mo}, \citenamefont {Zhou},
  \citenamefont {Yang}, \citenamefont {Kirchmann}, \citenamefont {Lu},
  \citenamefont {Moore}, \citenamefont {Hussain}, \citenamefont {Shen},
  \citenamefont {Qi},\ and\ \citenamefont {Sasagawa}}]{Chen}%
  \BibitemOpen
  \bibfield  {author} {\bibinfo {author} {\bibfnamefont {Y.~L.}\ \bibnamefont
  {Chen}}, \bibinfo {author} {\bibfnamefont {M.}~\bibnamefont {Kanou}},
  \bibinfo {author} {\bibfnamefont {Z.~K.}\ \bibnamefont {Liu}}, \bibinfo
  {author} {\bibfnamefont {H.~J.}\ \bibnamefont {Zhang}}, \bibinfo {author}
  {\bibfnamefont {J.~A.}\ \bibnamefont {Sobota}}, \bibinfo {author}
  {\bibfnamefont {D.}~\bibnamefont {Leuenberger}}, \bibinfo {author}
  {\bibfnamefont {S.~K.}\ \bibnamefont {Mo}}, \bibinfo {author} {\bibfnamefont
  {B.}~\bibnamefont {Zhou}}, \bibinfo {author} {\bibfnamefont {S.-L.}\
  \bibnamefont {Yang}}, \bibinfo {author} {\bibfnamefont {P.~S.}\ \bibnamefont
  {Kirchmann}}, \bibinfo {author} {\bibfnamefont {D.~H.}\ \bibnamefont {Lu}},
  \bibinfo {author} {\bibfnamefont {R.~G.}\ \bibnamefont {Moore}}, \bibinfo
  {author} {\bibfnamefont {Z.}~\bibnamefont {Hussain}}, \bibinfo {author}
  {\bibfnamefont {Z.~X.}\ \bibnamefont {Shen}}, \bibinfo {author}
  {\bibfnamefont {X.~L.}\ \bibnamefont {Qi}}, \ and\ \bibinfo {author}
  {\bibfnamefont {T.}~\bibnamefont {Sasagawa}},\ }\href@noop {} {\bibfield
  {journal} {\bibinfo  {journal} {Nat. Phys.}\ }\textbf {\bibinfo {volume}
  {9}},\ \bibinfo {pages} {704} (\bibinfo {year}
  {2013}{\natexlab{a}})}\BibitemShut {NoStop}%
\bibitem [{\citenamefont {Rashba}(2012)}]{Rashba12}%
  \BibitemOpen
  \bibfield  {author} {\bibinfo {author} {\bibfnamefont {E.~I.}\ \bibnamefont
  {Rashba}},\ }\href {http://link.aps.org/doi/10.1103/PhysRevB.86.125319}
  {\bibfield  {journal} {\bibinfo  {journal} {Phys. Rev. B}\ }\textbf {\bibinfo
  {volume} {86}},\ \bibinfo {pages} {125319} (\bibinfo {year}
  {2012})}\BibitemShut {NoStop}%
\bibitem [{\citenamefont {Datta}\ and\ \citenamefont {Das}(1990)}]{Datta89}%
  \BibitemOpen
  \bibfield  {author} {\bibinfo {author} {\bibfnamefont {S.}~\bibnamefont
  {Datta}}\ and\ \bibinfo {author} {\bibfnamefont {B.}~\bibnamefont {Das}},\
  }\href {\doibase 10.1063/1.102730} {\bibfield  {journal} {\bibinfo  {journal}
  {Appl. Phys. Lett.}\ }\textbf {\bibinfo {volume} {56}},\ \bibinfo {pages}
  {665} (\bibinfo {year} {1990})}\BibitemShut {NoStop}%
\bibitem [{\citenamefont {Koo}\ \emph {et~al.}(2009)\citenamefont {Koo},
  \citenamefont {Kwon}, \citenamefont {Eom}, \citenamefont {Chang},
  \citenamefont {Han},\ and\ \citenamefont {Johnson}}]{Koo09}%
  \BibitemOpen
  \bibfield  {author} {\bibinfo {author} {\bibfnamefont {H.~C.}\ \bibnamefont
  {Koo}}, \bibinfo {author} {\bibfnamefont {J.~H.}\ \bibnamefont {Kwon}},
  \bibinfo {author} {\bibfnamefont {J.}~\bibnamefont {Eom}}, \bibinfo {author}
  {\bibfnamefont {J.}~\bibnamefont {Chang}}, \bibinfo {author} {\bibfnamefont
  {S.~H.}\ \bibnamefont {Han}}, \ and\ \bibinfo {author} {\bibfnamefont
  {M.}~\bibnamefont {Johnson}},\ }\href {\doibase 10.1126/science.1173667}
  {\bibfield  {journal} {\bibinfo  {journal} {Science}\ }\textbf {\bibinfo
  {volume} {325}},\ \bibinfo {pages} {1515} (\bibinfo {year}
  {2009})}\BibitemShut {NoStop}%
\bibitem [{\citenamefont {Chen}\ \emph
  {et~al.}(2013{\natexlab{b}})\citenamefont {Chen}, \citenamefont {Kanou},
  \citenamefont {Liu}, \citenamefont {Zhang}, \citenamefont {Sobota},
  \citenamefont {Leuenberger}, \citenamefont {Mo}, \citenamefont {Zhou},
  \citenamefont {Yang}, \citenamefont {Kirchmann}, \citenamefont {Lu},
  \citenamefont {Moore}, \citenamefont {Hussain}, \citenamefont {Shen},
  \citenamefont {Qi},\ and\ \citenamefont {Sasagawa}}]{Sasagawa}%
  \BibitemOpen
  \bibfield  {author} {\bibinfo {author} {\bibfnamefont {Y.~L.}\ \bibnamefont
  {Chen}}, \bibinfo {author} {\bibfnamefont {M.}~\bibnamefont {Kanou}},
  \bibinfo {author} {\bibfnamefont {Z.~K.}\ \bibnamefont {Liu}}, \bibinfo
  {author} {\bibfnamefont {H.~J.}\ \bibnamefont {Zhang}}, \bibinfo {author}
  {\bibfnamefont {J.~A.}\ \bibnamefont {Sobota}}, \bibinfo {author}
  {\bibfnamefont {D.}~\bibnamefont {Leuenberger}}, \bibinfo {author}
  {\bibfnamefont {S.~K.}\ \bibnamefont {Mo}}, \bibinfo {author} {\bibfnamefont
  {B.}~\bibnamefont {Zhou}}, \bibinfo {author} {\bibfnamefont {S.-L.}\
  \bibnamefont {Yang}}, \bibinfo {author} {\bibfnamefont {P.~S.}\ \bibnamefont
  {Kirchmann}}, \bibinfo {author} {\bibfnamefont {D.~H.}\ \bibnamefont {Lu}},
  \bibinfo {author} {\bibfnamefont {R.~G.}\ \bibnamefont {Moore}}, \bibinfo
  {author} {\bibfnamefont {Z.}~\bibnamefont {Hussain}}, \bibinfo {author}
  {\bibfnamefont {Z.~X.}\ \bibnamefont {Shen}}, \bibinfo {author}
  {\bibfnamefont {X.~L.}\ \bibnamefont {Qi}}, \ and\ \bibinfo {author}
  {\bibfnamefont {T.}~\bibnamefont {Sasagawa}},\ }\href@noop {} {\bibfield
  {journal} {\bibinfo  {journal} {Nat. Phys.}\ }\textbf {\bibinfo {volume}
  {9}},\ \bibinfo {pages} {704} (\bibinfo {year}
  {2013}{\natexlab{b}})}\BibitemShut {NoStop}%
\bibitem [{\citenamefont {Souza}\ \emph {et~al.}(2001)\citenamefont {Souza},
  \citenamefont {Marzari},\ and\ \citenamefont {Vanderbilt}}]{TB_Wannier1}%
  \BibitemOpen
  \bibfield  {author} {\bibinfo {author} {\bibfnamefont {I.}~\bibnamefont
  {Souza}}, \bibinfo {author} {\bibfnamefont {N.}~\bibnamefont {Marzari}}, \
  and\ \bibinfo {author} {\bibfnamefont {D.}~\bibnamefont {Vanderbilt}},\
  }\href {\doibase 10.1103/PhysRevB.65.035109} {\bibfield  {journal} {\bibinfo
  {journal} {Phys. Rev. B}\ }\textbf {\bibinfo {volume} {65}},\ \bibinfo
  {pages} {035109} (\bibinfo {year} {2001})}\BibitemShut {NoStop}%
\bibitem [{\citenamefont {Wang}\ \emph {et~al.}(2006)\citenamefont {Wang},
  \citenamefont {Yates}, \citenamefont {Souza},\ and\ \citenamefont
  {Vanderbilt}}]{TB_Wannier2}%
  \BibitemOpen
  \bibfield  {author} {\bibinfo {author} {\bibfnamefont {X.}~\bibnamefont
  {Wang}}, \bibinfo {author} {\bibfnamefont {J.~R.}\ \bibnamefont {Yates}},
  \bibinfo {author} {\bibfnamefont {I.}~\bibnamefont {Souza}}, \ and\ \bibinfo
  {author} {\bibfnamefont {D.}~\bibnamefont {Vanderbilt}},\ }\href {\doibase
  10.1103/PhysRevB.74.195118} {\bibfield  {journal} {\bibinfo  {journal} {Phys.
  Rev. B}\ }\textbf {\bibinfo {volume} {74}},\ \bibinfo {pages} {195118}
  (\bibinfo {year} {2006})}\BibitemShut {NoStop}%
\bibitem [{\citenamefont {Mostofi}\ \emph {et~al.}(2008)\citenamefont
  {Mostofi}, \citenamefont {Yates}, \citenamefont {Lee}, \citenamefont {Souza},
  \citenamefont {Vanderbilt},\ and\ \citenamefont {Marzari}}]{TB_Wannier3}%
  \BibitemOpen
  \bibfield  {author} {\bibinfo {author} {\bibfnamefont {A.~A.}\ \bibnamefont
  {Mostofi}}, \bibinfo {author} {\bibfnamefont {J.~R.}\ \bibnamefont {Yates}},
  \bibinfo {author} {\bibfnamefont {Y.-S.}\ \bibnamefont {Lee}}, \bibinfo
  {author} {\bibfnamefont {I.}~\bibnamefont {Souza}}, \bibinfo {author}
  {\bibfnamefont {D.}~\bibnamefont {Vanderbilt}}, \ and\ \bibinfo {author}
  {\bibfnamefont {N.}~\bibnamefont {Marzari}},\ }\href {\doibase
  http://dx.doi.org/10.1016/j.cpc.2007.11.016} {\bibfield  {journal} {\bibinfo
  {journal} {Comput. Phys. Commun.}\ }\textbf {\bibinfo {volume} {178}},\
  \bibinfo {pages} {685 } (\bibinfo {year} {2008})}\BibitemShut {NoStop}%
\bibitem [{\citenamefont {Kune\v{s}}\ \emph {et~al.}(2010)\citenamefont
  {Kune\v{s}}, \citenamefont {Arita}, \citenamefont {Wissgott}, \citenamefont
  {Toschi}, \citenamefont {Ikeda},\ and\ \citenamefont {Held}}]{TB_Wannier4}%
  \BibitemOpen
  \bibfield  {author} {\bibinfo {author} {\bibfnamefont {J.}~\bibnamefont
  {Kune\v{s}}}, \bibinfo {author} {\bibfnamefont {R.}~\bibnamefont {Arita}},
  \bibinfo {author} {\bibfnamefont {P.}~\bibnamefont {Wissgott}}, \bibinfo
  {author} {\bibfnamefont {A.}~\bibnamefont {Toschi}}, \bibinfo {author}
  {\bibfnamefont {H.}~\bibnamefont {Ikeda}}, \ and\ \bibinfo {author}
  {\bibfnamefont {K.}~\bibnamefont {Held}},\ }\href {\doibase
  http://dx.doi.org/10.1016/j.cpc.2010.08.005} {\bibfield  {journal} {\bibinfo
  {journal} {Comput. Phys. Commun.}\ }\textbf {\bibinfo {volume} {181}},\
  \bibinfo {pages} {1888 } (\bibinfo {year} {2010})}\BibitemShut {NoStop}%
\bibitem [{\citenamefont {Savin}\ \emph {et~al.}(1997)\citenamefont {Savin},
  \citenamefont {Nesper}, \citenamefont {Wengert},\ and\ \citenamefont
  {F\"assler}}]{ELF}%
  \BibitemOpen
  \bibfield  {author} {\bibinfo {author} {\bibfnamefont {A.}~\bibnamefont
  {Savin}}, \bibinfo {author} {\bibfnamefont {R.}~\bibnamefont {Nesper}},
  \bibinfo {author} {\bibfnamefont {S.}~\bibnamefont {Wengert}}, \ and\
  \bibinfo {author} {\bibfnamefont {T.~F.}\ \bibnamefont {F\"assler}},\ }\href
  {\doibase 10.1002/anie.199718081} {\bibfield  {journal} {\bibinfo  {journal}
  {Angew. Chem. Int. Ed. Engl.}\ }\textbf {\bibinfo {volume} {36}},\ \bibinfo
  {pages} {1808} (\bibinfo {year} {1997})}\BibitemShut {NoStop}%
\bibitem [{\citenamefont {Giuliani}\ and\ \citenamefont
  {Vignale}(2005)}]{Giuliani}%
  \BibitemOpen
  \bibfield  {author} {\bibinfo {author} {\bibfnamefont {G.~F.}\ \bibnamefont
  {Giuliani}}\ and\ \bibinfo {author} {\bibfnamefont {G.}~\bibnamefont
  {Vignale}},\ }\href@noop {} {\emph {\bibinfo {title} {{\itshape Quantum
  theory of the electron liquid}}}}\ (\bibinfo  {publisher} {Cambridge
  University Press},\ \bibinfo {address} {Cambridge},\ \bibinfo {year}
  {2005})\BibitemShut {NoStop}%
\bibitem [{\citenamefont {Starke}\ and\ \citenamefont {Schober}(2015)}]{Refr}%
  \BibitemOpen
  \bibfield  {author} {\bibinfo {author} {\bibfnamefont {R.}~\bibnamefont
  {Starke}}\ and\ \bibinfo {author} {\bibfnamefont {G.~A.~H.}\ \bibnamefont
  {Schober}},\ }\href@noop {} {}\bibinfo {howpublished} {arXiv:1510.03404
  [cond-mat.mtrl-sci]} (\bibinfo {year} {2015})\BibitemShut {NoStop}%
\bibitem [{ELK()}]{ELK}%
  \BibitemOpen
  \href@noop {} {\emph {\bibinfo {title} {{LAPW code: ELK}}}},\ \bibinfo {note}
  {\url{http://elk.sourceforge.net}}\BibitemShut {NoStop}%
\bibitem [{\citenamefont {Andersen}(1975)}]{LAPW1}%
  \BibitemOpen
  \bibfield  {author} {\bibinfo {author} {\bibfnamefont {O.~K.}\ \bibnamefont
  {Andersen}},\ }\href@noop {} {\bibfield  {journal} {\bibinfo  {journal}
  {Phys. Rev. B}\ }\textbf {\bibinfo {volume} {12}},\ \bibinfo {pages} {3060}
  (\bibinfo {year} {1975})}\BibitemShut {NoStop}%
\bibitem [{\citenamefont {Koelling}\ and\ \citenamefont
  {Arbman}(1975)}]{LAPW2}%
  \BibitemOpen
  \bibfield  {author} {\bibinfo {author} {\bibfnamefont {D.~D.}\ \bibnamefont
  {Koelling}}\ and\ \bibinfo {author} {\bibfnamefont {G.~O.}\ \bibnamefont
  {Arbman}},\ }\href {http://stacks.iop.org/0305-4608/5/i=11/a=016} {\bibfield
  {journal} {\bibinfo  {journal} {J. Phys. F: Met. Phys.}\ }\textbf {\bibinfo
  {volume} {5}},\ \bibinfo {pages} {2041} (\bibinfo {year} {1975})}\BibitemShut
  {NoStop}%
\bibitem [{\citenamefont {{Sj\"ostedt}}\ \emph {et~al.}(2000)\citenamefont
  {{Sj\"ostedt}}, \citenamefont {{Nordstr\"om}},\ and\ \citenamefont
  {Singh}}]{LAPW3}%
  \BibitemOpen
  \bibfield  {author} {\bibinfo {author} {\bibfnamefont {E.}~\bibnamefont
  {{Sj\"ostedt}}}, \bibinfo {author} {\bibfnamefont {L.}~\bibnamefont
  {{Nordstr\"om}}}, \ and\ \bibinfo {author} {\bibfnamefont {D.}~\bibnamefont
  {Singh}},\ }\href {\doibase http://dx.doi.org/10.1016/S0038-1098(99)00577-3}
  {\bibfield  {journal} {\bibinfo  {journal} {Solid State Commun.}\ }\textbf
  {\bibinfo {volume} {114}},\ \bibinfo {pages} {15 } (\bibinfo {year}
  {2000})}\BibitemShut {NoStop}%
\bibitem [{\citenamefont {Singh}\ and\ \citenamefont
  {Nordstrom}(2006)}]{LAPW4}%
  \BibitemOpen
  \bibfield  {author} {\bibinfo {author} {\bibfnamefont {D.~J.}\ \bibnamefont
  {Singh}}\ and\ \bibinfo {author} {\bibfnamefont {L.}~\bibnamefont
  {Nordstrom}},\ }\href@noop {} {\emph {\bibinfo {title} {Planewaves,
  Pseudopotentials, and the LAPW method}}}\ (\bibinfo  {publisher} {Springer
  Science \& Business Media},\ \bibinfo {year} {2006})\BibitemShut {NoStop}%
\bibitem [{\citenamefont {Perdew}\ \emph {et~al.}(1996)\citenamefont {Perdew},
  \citenamefont {Burke},\ and\ \citenamefont {Ernzerhof}}]{pbe}%
  \BibitemOpen
  \bibfield  {author} {\bibinfo {author} {\bibfnamefont {J.~P.}\ \bibnamefont
  {Perdew}}, \bibinfo {author} {\bibfnamefont {K.}~\bibnamefont {Burke}}, \
  and\ \bibinfo {author} {\bibfnamefont {M.}~\bibnamefont {Ernzerhof}},\
  }\href@noop {} {\bibfield  {journal} {\bibinfo  {journal} {Phys. Rev. Lett.}\
  }\textbf {\bibinfo {volume} {77}},\ \bibinfo {pages} {3865} (\bibinfo {year}
  {1996})}\BibitemShut {NoStop}%
\bibitem [{\citenamefont {R{\"o}ssler}\ \emph {et~al.}(1989)\citenamefont
  {R{\"o}ssler}, \citenamefont {Malcher},\ and\ \citenamefont
  {Lommer}}]{SpinSplitting1}%
  \BibitemOpen
  \bibfield  {author} {\bibinfo {author} {\bibfnamefont {U.}~\bibnamefont
  {R{\"o}ssler}}, \bibinfo {author} {\bibfnamefont {F.}~\bibnamefont
  {Malcher}}, \ and\ \bibinfo {author} {\bibfnamefont {G.}~\bibnamefont
  {Lommer}},\ }in\ \href {\doibase 10.1007/978-3-642-83810-1_58} {\emph
  {\bibinfo {booktitle} {High Magnetic Fields in Semiconductor Physics {II:}
  Transport and Optics, Proceedings of the International Conference,
  W{\"u}rzburg, Fed. Rep. of Germany, August 22--26, 1988}}},\ \bibinfo
  {editor} {edited by\ \bibinfo {editor} {\bibfnamefont {G.}~\bibnamefont
  {Landwehr}}}\ (\bibinfo  {publisher} {Springer Berlin Heidelberg},\ \bibinfo
  {address} {Berlin, Heidelberg},\ \bibinfo {year} {1989})\ pp.\ \bibinfo
  {pages} {376--385}\BibitemShut {NoStop}%
\bibitem [{\citenamefont {Winkler}\ \emph {et~al.}(2001)\citenamefont
  {Winkler}, \citenamefont {Papadakis}, \citenamefont {De~Poortere},\ and\
  \citenamefont {Shayegan}}]{SpinSplitting2}%
  \BibitemOpen
  \bibfield  {author} {\bibinfo {author} {\bibfnamefont {R.}~\bibnamefont
  {Winkler}}, \bibinfo {author} {\bibfnamefont {S.~J.}\ \bibnamefont
  {Papadakis}}, \bibinfo {author} {\bibfnamefont {E.~P.}\ \bibnamefont
  {De~Poortere}}, \ and\ \bibinfo {author} {\bibfnamefont {M.}~\bibnamefont
  {Shayegan}},\ }in\ \href {\doibase 10.1007/3-540-44946-9_18} {\emph {\bibinfo
  {booktitle} {Advances in Solid State Physics}}},\ \bibinfo {editor} {edited
  by\ \bibinfo {editor} {\bibfnamefont {B.}~\bibnamefont {Kramer}}}\ (\bibinfo
  {publisher} {Springer},\ \bibinfo {address} {Berlin, Heidelberg},\ \bibinfo
  {year} {2001})\ pp.\ \bibinfo {pages} {211--223}\BibitemShut {NoStop}%
\bibitem [{\citenamefont {Shevelkov}\ \emph {et~al.}()\citenamefont
  {Shevelkov}, \citenamefont {Dikarev}, \citenamefont {Shpanchenko},\ and\
  \citenamefont {Popovkin}}]{XRAY_STRUCTURES}%
  \BibitemOpen
  \bibfield  {author} {\bibinfo {author} {\bibfnamefont {A.~V.}\ \bibnamefont
  {Shevelkov}}, \bibinfo {author} {\bibfnamefont {E.~V.}\ \bibnamefont
  {Dikarev}}, \bibinfo {author} {\bibfnamefont {R.~V.}\ \bibnamefont
  {Shpanchenko}}, \ and\ \bibinfo {author} {\bibfnamefont {B.~A.}\ \bibnamefont
  {Popovkin}},\ }\href {\doibase 10.1006/jssc.1995.1058} {\ \textbf {\bibinfo
  {volume} {114}},\ \bibinfo {pages} {379}}\BibitemShut {NoStop}%
\bibitem [{\citenamefont {Rodr\'iguez-Carvajal}(1993)}]{Fullprof}%
  \BibitemOpen
  \bibfield  {author} {\bibinfo {author} {\bibfnamefont {J.}~\bibnamefont
  {Rodr\'iguez-Carvajal}},\ }\href {\doibase
  http://dx.doi.org/10.1016/0921-4526(93)90108-I} {\bibfield  {journal}
  {\bibinfo  {journal} {Physica B: Condens. Matter}\ }\textbf {\bibinfo
  {volume} {192}},\ \bibinfo {pages} {55 } (\bibinfo {year}
  {1993})}\BibitemShut {NoStop}%
\bibitem [{\citenamefont {Momma}\ and\ \citenamefont
  {Izumi}(2011)}]{momma_vesta_2011}%
  \BibitemOpen
  \bibfield  {author} {\bibinfo {author} {\bibfnamefont {K.}~\bibnamefont
  {Momma}}\ and\ \bibinfo {author} {\bibfnamefont {F.}~\bibnamefont {Izumi}},\
  }\href@noop {} {\bibfield  {journal} {\bibinfo  {journal} {J. Appl.
  Crystallogr.}\ }\textbf {\bibinfo {volume} {44}},\ \bibinfo {pages} {1272}
  (\bibinfo {year} {2011})}\BibitemShut {NoStop}%
\bibitem [{\citenamefont {Rusinov}\ \emph {et~al.}(2013)\citenamefont
  {Rusinov}, \citenamefont {Nechaev}, \citenamefont {Eremeev}, \citenamefont
  {Friedrich}, \citenamefont {Bl\"ugel},\ and\ \citenamefont
  {Chulkov}}]{Rusinov13}%
  \BibitemOpen
  \bibfield  {author} {\bibinfo {author} {\bibfnamefont {I.~P.}\ \bibnamefont
  {Rusinov}}, \bibinfo {author} {\bibfnamefont {I.~A.}\ \bibnamefont
  {Nechaev}}, \bibinfo {author} {\bibfnamefont {S.~V.}\ \bibnamefont
  {Eremeev}}, \bibinfo {author} {\bibfnamefont {C.}~\bibnamefont {Friedrich}},
  \bibinfo {author} {\bibfnamefont {S.}~\bibnamefont {Bl\"ugel}}, \ and\
  \bibinfo {author} {\bibfnamefont {E.~V.}\ \bibnamefont {Chulkov}},\ }\href
  {\doibase 10.1103/PhysRevB.87.205103} {\bibfield  {journal} {\bibinfo
  {journal} {Phys. Rev. B}\ }\textbf {\bibinfo {volume} {87}},\ \bibinfo
  {pages} {205103} (\bibinfo {year} {2013})}\BibitemShut {NoStop}%
\bibitem [{\citenamefont {Secuk}\ and\ \citenamefont
  {Akkus}(2016)}]{BiTeBr_optics}%
  \BibitemOpen
  \bibfield  {author} {\bibinfo {author} {\bibfnamefont {M.~N.}\ \bibnamefont
  {Secuk}}\ and\ \bibinfo {author} {\bibfnamefont {H.}~\bibnamefont {Akkus}},\
  }in\ \href@noop {} {\emph {\bibinfo {booktitle} {JPCS}}},\ Vol.\ \bibinfo
  {volume} {707}\ (\bibinfo {organization} {IOP Publishing},\ \bibinfo {year}
  {2016})\ p.\ \bibinfo {pages} {012017}\BibitemShut {NoStop}%
\bibitem [{\citenamefont {Guo}\ and\ \citenamefont {Wang}(2016)}]{Guo16}%
  \BibitemOpen
  \bibfield  {author} {\bibinfo {author} {\bibfnamefont {S.-D.}\ \bibnamefont
  {Guo}}\ and\ \bibinfo {author} {\bibfnamefont {J.-L.}\ \bibnamefont {Wang}},\
  }\href {http://stacks.iop.org/0022-3727/49/i=21/a=215107} {\bibfield
  {journal} {\bibinfo  {journal} {J. Phys. D: Appl. Phys.}\ }\textbf {\bibinfo
  {volume} {49}},\ \bibinfo {pages} {215107} (\bibinfo {year}
  {2016})}\BibitemShut {NoStop}%
\bibitem [{\citenamefont {G\"uler-K\ifmmode \imath \else \i \fi{}l\ifmmode
  \imath \else \i \fi{}\ifmmode~\mbox{\c{c}}\else \c{c}\fi{}}\ and\
  \citenamefont {K\ifmmode \imath \else \i \fi{}l\ifmmode \imath \else \i
  \fi{}\ifmmode~\mbox{\c{c}}\else \c{c}\fi{}}(2015)}]{Kilic}%
  \BibitemOpen
  \bibfield  {author} {\bibinfo {author} {\bibfnamefont {S.}~\bibnamefont
  {G\"uler-K\ifmmode \imath \else \i \fi{}l\ifmmode \imath \else \i
  \fi{}\ifmmode~\mbox{\c{c}}\else \c{c}\fi{}}}\ and\ \bibinfo {author}
  {\bibfnamefont {{\c{C}}.}~\bibnamefont {K\ifmmode \imath \else \i
  \fi{}l\ifmmode \imath \else \i \fi{}\ifmmode~\mbox{\c{c}}\else \c{c}\fi{}}},\
  }\href {\doibase 10.1103/PhysRevB.91.245204} {\bibfield  {journal} {\bibinfo
  {journal} {Phys. Rev. B}\ }\textbf {\bibinfo {volume} {91}},\ \bibinfo
  {pages} {245204} (\bibinfo {year} {2015})}\BibitemShut {NoStop}%
\end{thebibliography}%

\end{document}